\documentclass[a4paper,11pt]{article}
\usepackage{graphicx,graphics,xcolor}
\usepackage{epsfig}
\usepackage{amsmath}
\usepackage{amssymb}
\usepackage{mathtools}
\usepackage[top=1cm, bottom=1.5cm, left=1.2cm, right=1.2cm]{geometry}
\usepackage{microtype}
\usepackage{lmodern}
\usepackage[utf8]{inputenc}
\usepackage{float}
\usepackage{hyperref}
\usepackage{bera}
\newcommand{\ds}{\displaystyle }

\newcommand{\beq}{\begin{equation} }
\newcommand{\eeq}{\end{equation}}
\begin{document}

\title{From random walks to epidemic spreading: Compartment model with mortality for vector transmitted diseases\footnote{PROCEEDINGS ICDEA 2024} }
\author{
T\'eo Granger$^1$, Thomas M. Michelitsch$^1$\footnote{Corresponding, \, E-mail:\, thomas.michelitsch@sorbonne-universite.fr}, Bernard A. Collet$^1$  \\
Michael Bestehorn$^2$, Alejandro P. Riascos$^3$ \\[3ex] 
$^1$Sorbonne Université, CNRS, Institut Jean Le Rond d’Alembert, F-75005 Paris, France
\\ \\[1ex]
$^2$Brandenburgische Technische Universit\"at Cottbus-Senftenberg \\
Institut f\"ur Physik \\ Erich-Weinert-Stra{\ss}e 1, 03046 Cottbus, Germany
\\ \\[1ex]
$^3$Departamento de Física, Universidad Nacional de Colombia, 
Bogot\'a, Colombia  
}
\maketitle
\begin{abstract}
We propose a compartmental model for vector-transmitted diseases, such as Malaria and Dengue, spreading over complex networks. Individuals are represented by independent random walkers and vectors by infected nodes. Both walkers and nodes can be susceptible (S) or infected (I). Infected walkers may die (entering the dead compartment D), while infected nodes remain alive. Susceptible walkers can be infected by visiting infected nodes, and susceptible nodes by visits from infected walkers. We derive explicit expressions for the basic reproduction numbers $R_0 $ (without mortality) and $R_M$ (with mortality), proving that  $ R_M < R_0$. When $R_M, R_0 > 1$, the healthy state is unstable, and for zero mortality, an endemic equilibrium emerges. We also study the effects of confinement measures. Simulations align well with mean-field predictions on strongly connected graphs but deviate for weakly connected networks. Our model has various interdisciplinary applications which include the modeling of chemical reaction kinetics, contaminant spread, and wildfire propagation.
\\[1ex]
\end{abstract}
\newpage
\tableofcontents
\section{Introduction}
\label{Intro}

Since the breakout of the COVID-19 pandemics, epidemic models have attracted considerable attention. A popular way to tackle epidemic spreading are so-called compartmental models, where the individuals of a population are divided 
according to  their states of health. 
The first compartmental model traces back to the seminal work of Kermack and McKendrick \cite{KermackMcKendrick1927}, where
an individual is in one of the states (compartments)
susceptible (to infection) - S, infected and infectious - I, recovered (immune) - R.
Standard models of the SIR class are able to describe several features of certain infectious diseases, which include 
mumps, measles, rubella and others. However, standard SIR models do not exhibit persistent oscillatory behaviors or spontaneous outbursts, features which are often observed in the evolution of epidemics. One of the first works able to capture periodic behavior goes back to the model of Soper \cite{Soper1929}.

The classical SIR model has been generalized in many directions \cite{LiuHeathcote1987,Li-etal1999,Anderson1992,Martcheva2015} and consult \cite{Harris2023} and references therein for
models related to the context of COVID-19 pandemics.
In order to relate macroscopic compartment models to microscopic dynamics, epidemic spreading  
has been studied in random graphs with emphasis on the complex interplay of the network topology and spreading features
\cite{Satoras-Vespignani-etal2015,Pastor-SatorrasVespignani2001,OkabeShudo2021}.

A further class are stochastic compartmental models combined with random walk approaches
\cite{Barabasi2016,BarabasiAlbert1999,Barrat-etal2008,Ross1996,fractional_book_MiRia2019,RiascosSanders2021,BesMi-etal2021,vanKampen1981,VanMiegem2014} and non-exponentially distributed compartmental sojourn times leading to non-Markovian models were considered recently \cite{BasnakovSandev-etal2020,BesMiRias2022,Granger-et-al2023,BesMi2023}.

\begin{figure}[H]
\centerline{\includegraphics[width=0.7\textwidth]{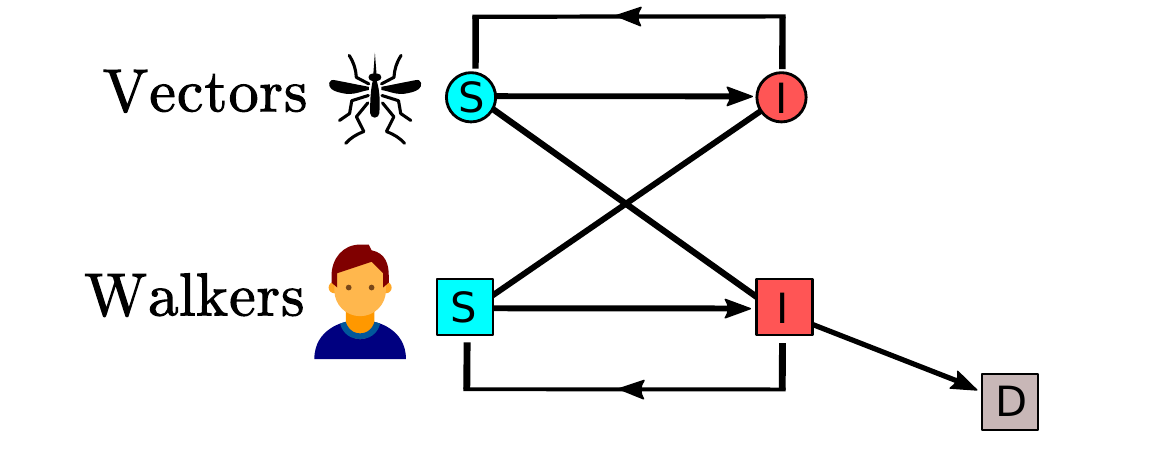}}
\vspace{-3mm}
\caption{Vector-borne transmission pathway with mortality of infected walkers.}
\label{Fig_model}
\end{figure}

In the present paper, our aim is to study the spreading of the class of vector transmitted diseases in a large population of individuals (random walkers) moving
on complex graphs similar to human mobility patterns in complex environments such as cities, street-, and transportation networks. The walkers are mimicking the individuals of the population, and the nodes represent the vectors.
Susceptible walkers may be infected by visiting infected nodes, and susceptible nodes by visits from infected walkers. 
We also account for the mortality of infected individuals (random walkers) and assume absence of mortality for the infected nodes (vectors). The latter assumption is based on the observation that the vectors (for instance mosquitos) do not fall ill during their infection, and therefore may not die from it. 
A schematic representation of the vector-borne transmission pathway is shown in Fig. \ref{Fig_model}.
The paper has the following structure. In Sect. \ref{SIS-motality} we establish a mean field model and obtain
macroscopic evolution equations in stochastic and explicit forms. We derive the basic reproduction number $R_M$ in presence of mortality and show that $R_M \leq R_0$, i.e. the basic reproduction number $R_0$ without mortality is un upper bound. We show by linear stability analysis that the healthy state is unstable for $R_R,R_M >1$ and stable for $R_R,R_M < 1$. 
For zero mortality ($R_M = R_0$) a globally stable endemic equilibrium  exists for $R_0>1$ independent of the initial conditions.
In Sect. \ref{RW_simulations} we compare the outcomes of the mean field model with random walk simulations,
where we focus on Watts-Strogatz (WS) and Barabási–Albert (BA) graphs. At the end of the paper, we investigate for zero mortality the effect of confinement measures on the spreading of the disease and the 
basic reproduction number in the small world BA graph and compare it with the case of the large world WS graph. 
For more detailed demonstrations and discussions, we refer to our recent article \cite{SISI_Entropy_2024}.

\section{Stochastic compartmental model with mortality}
\label{SIS-motality}
\subsection{Mean field approach}
\label{mean_field}

Here, we recall briefly the essentials of our macroscopic mean field model for infectious diseases
which are transmitted via vectors. 
Vector transmitted diseases include Dengue, Malaria (transmission vectors are mosquitos)
and Pestilence (vectors are fleas), and others \cite{Whitehead2007}.

The vectors of the disease are represented by infected nodes of the network. This assumption is based on the observation
that in the cases of Malaria and Dengue, the transmitting mosquitoes live in specific areas such as swamps and forests.
Each walker performs independent steps along edges connecting nodes of the network. Both walkers and nodes are in one of the states (compartments): S -- susceptible or I -- infected. In addition, walkers can die during their infection by performing a transition from the I to the dead (D) compartment. Infected nodes representing the vectors never die. This feature accounts for the observation that infected mosquitos are not ill and therefore do not die from their infection.

It has to be pointed out that for some real-world diseases mortality of vectors may be well non-negligible.
However, we focus on the class of diseases, where mortality of vectors is not significant as they do not exhibit symptoms of illness during their infection.
We also do not account for natural birth and death processes. 
This means we consider diseases, where the spreading happens on time-scales for which the demographic change of the population numbers can be neglected.
There might exist "slowly" spreading real-world diseases, where natural birth and death processes should be taken into account.

We introduce $Z_S(t), Z_I(t)$ ($N_S(t), N_I(t)$) indicating the number of walkers (nodes) in compartments S and I, and $Z_D(t)$ the non-decreasing number of walkers (in compartment D) died during their infection up to time $t$. 
We consider the total number of walkers $Z = Z(0) =  Z_I(t)+ Z_S(t) + Z_D(t)$ 
and of nodes $N = N_I(t)+N_S(t)$ to be constant, independent of time. We account only for mortality of walkers during their infections. Natural birth and death processes are not considered in the present study.

We assume that at time instant $t=0$ the first spontaneous infections occur:  $Z_I(0) \ll Z$ walkers and $N_I(0) \ll N$ nodes and no dead walkers $Z_D(0)=0$.
It is convenient to introduce the compartmental fractions 
$S_w(t) = \frac{Z_S(t)}{Z}$, $J_w(t)=\frac{Z_I(t)}{Z}$, $d_w(t)= \frac{Z_D(t)}{Z}$ for the walkers (normalized with respect to the time-independent total number $Z$ of walkers which includes the number $Z_D(t)$ of dead walkers) with $S_w(t)+J_w(t)+d_w(t)=1$, and $S_n(t)= \frac{N_S(t)}{N}$, $J_n(t)=\frac{N_I(t)}{N}$ with $S_n(t)+J_n(t)=1$.
\\[1ex]
A list of the used symbols and quantities is given hereafter and detailed explanations are given in the text.
\begin{center}
\begin{tabular}{c l}
\label{tab1} 
 $R_0$ :& basic reproduction number without mortality\\ 
 $R_M$ :& basic reproduction number with mortality\\  
 $S_w$ : & fraction of susceptible walkers \\
 $J_w$ : & fraction of infected walkers \\
 $d_w$ : & fraction of dead walkers \\
 $S_n$ : & fraction of susceptible nodes \\
 $J_n$ : & fraction of infected nodes \\
 ${\cal A}_w$ : & infection rate of walkers \\
  ${\cal A}_n$ : & infection rate of nodes \\
  $\beta_w$ : & infection rate parameter of walkers \\
  $\beta_n$ : & infection rate parameter of nodes \\
  $t_I^w$ :  & random infection time span of walker without mortality \\
   $t_I^n$ :  & random infection time span of nodes \\
   $t_M$ : & random survival time span of infected walker \\
   $t_{I M}^w$ : & random infection time span of infected mortal walker \\
   $b_r$ : & defective recovery PDF of walker \\
   $b_d$ : & defective mortality PDF of walker \\
   $\Phi_I^w$ : & persistence probability of walkers' infection without mortality\\
    $\Phi_I^n$ : & persistence probability of nodes' infection \\
    $\Phi_M$ : & survival probability of infected walker for infinitely long infection time span \\ 
    ${\cal K}_{I,M}^w$ : & end of infection PDF of mortal walker \\
    ${\cal K}_I^w$ : & recovery PDF of walker without mortality\\
    ${\cal K}_I^n$ : & recovery PDF of node \\
    ${\cal R}$ : & cumulative recovery probability of walker \\
    ${\cal D}$ : & cumulative mortality probability of walker \\
    $\Theta(\ldots)$ : & Heaviside unit-step function \\
    $\delta(\ldots)$ : & Dirac's $\delta$-distribution  \\
    $\big\langle (\ldots) \big\rangle$ : & average over the random variables contained in $(\ldots)$ \\
    $ {\hat f}$ : & Laplace transform of $f$ .
\end{tabular}
\end{center}
We denote with ${\cal A}_w(t), {\cal A}_n(t)$ the infection 
rates (rates of transitions S $\to$ I) of walkers and nodes, respectively.
We assume the simple bi-linear forms inspired from mass-action laws
\beq
\label{infections_rate}
\begin{array}{clr}
\ds {\cal A}_w(t)&  = \ds {\cal A}_w[S_w(t), J_n(t)]   =  \beta_w S_w(t)J_n(t) & \\ \\
\ds {\cal A}_n(t) & = \ds {\cal A}_n[S_n(t), J_w(t)]   =  \beta_n S_n(t)J_w(t) &
\end{array}
\eeq
with time-independent constant rate parameters $\beta_w, \beta_n >0$. Eqs. (\ref{infections_rate}) correspond to a predator-prey mechanism, where susceptible walkers (nodes) are the prey of the predator, the infected nodes (walkers).
${\cal A}_w(t)$ indicates the infection rate of walkers, the susceptible walkers $S_w$ may be infected with a certain probability by (visiting) infected nodes $J_n(t)$. $ {\cal A}_n(t)$ stands for the infection rate of nodes, indicating that susceptible nodes $S_n(t)$ may be 
infected with a certain probability by visits of infected walkers $J_w(t)$. 
No direct transmissions among walkers and among nodes can happen. See again Fig. \ref{Fig_model} for a schematic representation of the transmission pathway.
Infections occur with specific transmission probabilities $p_{w,n}$ which determine the transmission rate constants $\beta_{w,n}$. The latter contain additional topological information when considering the spreading on networks.

We assume that the infection time spans $t_I^w, t_I^n >0$ without mortality (waiting times in compartment I)
of walkers and nodes, respectively, are independent random variables drawn from specific probability density functions (PDFs).
Further, as the only type of mortality, we allow only infected walkers to die. In other words, we admit walkers to die only during the phases of their infection. 
To this end, we introduce the independent random variable $t_M>0$ measuring the survival time span of an infected walker. Both the infection and life time spans $t_I^w, t_M$ are counted from 
the moment of the infection event.
An infected walker survives the disease only if $t_M > t_I^w$ and dies for $t_M < t_I^w$. 
With these considerations, we are able to give a stochastic formulation of the evolution equations for the compartmental fractions
\beq
\label{evoleqs}
\begin{array}{clr}
\ds \frac{d}{dt}S_w(t) & = \ds   - {\cal A}_w(t) + 
\left\langle {\cal A}_w(t-t_I^w)
\Theta(t_M-t_I^w)\right\rangle  + J_w(0)\langle \delta(t-t_I^w)\Theta(t_M-t_I^w) \rangle & 
 \\ \\ 
\ds \frac{d}{dt} J_w(t) & = \ds  {\cal A}_w(t)    - \left\langle \,  {\cal A}_w(t-t_I^w) \Theta(t_M-t_I^w)\, \right\rangle  
 -J_w(0)\left\langle \,\delta(t-t_I^w)\Theta(t_M-t_I^w) \, \right\rangle    -\frac{d}{dt}d_w(t)  &   \\ \\
\ds \frac{d}{dt} d_w(t) = & \ds 
\left\langle \, {\cal A}_w(t-t_M) \Theta(t_I^w-t_M)\, \right\rangle +  J_w(0) \langle \delta(t-t_M)\Theta(t_I^w-t_M) \rangle
   &   \\ \\
\ds \frac{d}{dt}S_n(t) & = \ds - {\cal A}_n(t) +  \left\langle {\cal A}_n(t-t_I^n) \right\rangle + J_n(0)\langle \delta(t-t_I^n)\rangle  & \\ \\
\ds \frac{d}{dt}J_n(t) & =  \ds  -\frac{d}{dt}S_n(t) . &
\end{array}
\eeq
$ \frac{d}{dt}d_w(t)$ stands for the (non-negative) mortality rate of infected walkers (coinciding with the total mortality rate of walkers).
The symbol $\left\langle ..\right\rangle$ indicates averaging over the contained set of independent random variables $t_I^{w,n},t_M$. With
$\Theta(..)$ we denote the Heaviside unit-step function and with $\delta(..)$ the Dirac's $\delta$-distribution. 
The system is coupled by the infection rates (\ref{infections_rate}).
In Eqs. (\ref{evoleqs}) is taken into account that infected walkers may die during the time span of their infection, i.e. when $t_I^w > t_M$ (thus $\Theta(t_I^w-t_M) =1$)
and otherwise (for $t_I^w < t_M$) an infected walker recovers with a transition I to S when the time span $t_I^w$ of its infection has elapsed. Letting $t_M \to \infty$ recovers the model without mortality.
The last two equations of (\ref{evoleqs}) take into account that there is no mortality for
infected nodes. The contributions involving $\delta$-distributions account for the transitions and mortality of the initially infected walkers and of initially infected nodes without mortality.
We start the observation at $t=0$ and assume the initial conditions $S_{w,n}(0)=1-J_{w,n}(0)$ with $J_{w,n}(0) > 0$ with presence of a few infected walkers and/or nodes in a large susceptible population without dead walkers $d_w(0)=0$. 

We can average (\ref{evoleqs}) over the independent random variables $T_i= \{t_I^w, t_I^n,t_M\}$ using
\beq
\label{averaging_general}
\left\langle f(T_1,T_2,T_3) \right\rangle = \int_0^{\infty}\int_0^{\infty}\int_0^{\infty}
K_1(\tau_1)K_2(\tau_2)K_3(\tau_3)f(\tau_1,\tau_2,\tau_3){\rm d}\tau_1{\rm d}\tau_2{\rm d}\tau_3
\eeq
with their respective PDFs $K_i(\tau)=\{ K_{I}^{w}(\tau),K_{I}^{n}(\tau), K_M(\tau) \}$
and accounting for causality of the infection rates and PDFs with ${\cal A}_{w,n}(t), K_i(t) =0$ for $t<0$ taking into account that no infections occur prior to the outbreak of the disease at $t=0$
and for the positivity of $T_i= \{t_I^w, t_I^n,t_M\}$. An important special case arises for $f(T_1,T_2,T_3) = g_1(T_1) g_2(T_2) g_3(T_3) $, 
where from the independence of the $T_i$ follows the convenient property
\beq
\label{indep_facorize}
\left\langle f(T_1,T_2,T_3) \right\rangle = \left\langle g_1(T_1) \right\rangle \left\langle g_2(T_2) \right\rangle \left\langle g_3(T_3) \right\rangle 
\eeq
which we will use throughout this paper. These ingredients take us to
\beq
\label{causal_average}
\left\langle {\cal A}_{w,n}(t-T_i) \right\rangle = \int_0^t {\cal A}_{w,n}(t-\tau)K_i(\tau){\rm d}\tau .
\eeq
Useful are the persistence probabilities
\beq
\label{persistence}
\Phi_i(t) = Prob(T_i>t) = \left\langle \Theta(T_i-t) \right\rangle = \int_t^{\infty} K_i(\tau) {\rm d}\tau
\eeq
with $\Phi_i(0)=1$ reflecting normalization of the PDFs $K_i(\tau)$. Further worthy of mention is the relation
$\langle \delta(t-T_i) \rangle =K_i(t)$ from which follows that $\frac{d}{dt}\Phi_i(t) = -K_i(t)$, 
and consult \cite{SISI_Entropy_2024} for further details.
With these considerations, the explicit form of the averaged system (\ref{evoleqs}) reads
\beq
\label{evoleqsB}
\begin{array}{clr}
\ds \frac{d}{dt}S_w(t) & = \ds - {\cal A}_w(t) +   \int_0^t {\cal A}_w(t-\tau)K_I^w(\tau)\Phi_M(\tau) {\rm d}\tau + 
J_w(0)K_I^w(t)\Phi_M(t)   & 
 \\ \\
\ds \frac{d}{dt}J_w(t) & = \ds {\cal A}_w(t) - \int_0^t  {\cal A}_w(t-\tau) {\cal K}_{I,M}^w(\tau){\rm d}\tau 
- J_w(0) {\cal K}_{I,M}^w(t)
\ds     &  \\ \\
\frac{d}{dt}d_w(t) & =  \ds  \int_0^t{\cal A}_w(t-\tau) K_M(\tau)\Phi_I^w(\tau){\rm d}\tau + J_w(0)K_M(t)\Phi_I^w(t)   & \\ \\
\ds \frac{d}{dt}S_n(t) & = \ds - {\cal A}_n(t) +  
\int_0^t {\cal A}_n(t-\tau)K_I^n(\tau) {\rm d}\tau + J_n(0)K_I^n(t) & \\ \\
\ds \frac{d}{dt}J_n(t) & =  \ds - \frac{d}{dt}S_n(t). & \\ & &
\end{array}
\eeq
In the second equation of this system comes into play the PDF ${\cal K}_{I,M}^w(t) = K_I^w(t)\Phi_M(t)+K_M(t)\Phi_I^w(t)$ 
capturing all exits of walkers from compartment I, in which
the term 
\beq
\label{su_ra}
b_r(t)= \langle \delta(t-t_I^w)\Theta(t_M-t_I^w) \rangle = K_I^w(t)\Phi_M(t)
\eeq
describes the recovery rate of walkers that survived the disease, and 
\beq
\label{mo_ra}
b_d(t)= \langle \delta(t-t_M)\Theta(t_I^w-t_M) \rangle  = K_M(t)\Phi_I^w(t)
\eeq
the mortality rate, both referring to $J_w(0)=1$. The quantities $b_r(t),b_d(t)$ represent `defective' PDFs and are not 
properly normalized \cite{donofrio2024}.
Observe that
\beq
\label{KMIW}
{\cal K}_{I,M}^w(t) = b_r(t)+b_d(t) =-\frac{d}{dt} [\langle \Theta(t_M-t)\Theta(t_I^w-t) \rangle ]  = -\frac{d}{dt}[\Phi_I^w(t)\Phi_M(t)] ,
\eeq
where proper normalization of ${\cal K}_{I,M}^w(t)$ can be readily retrieved from 
\beq
\label{auxiliary_rel}
\int_0^{\infty}{\cal K}_{I,M}^w(t) {\rm d}t = - \int_0^{\infty} \frac{d}{dt}[\Phi_I^w(t)\Phi_M(t)] {\rm d}t =  \Phi_I^w(0)\Phi_M(0) =1 ,
\eeq
where $\Phi_I^w(\infty), \Phi_M(\infty) = 0$. 
Moreover, one has
\beq
\label{persist_infect}
\left\langle \Theta(t_M-t)\Theta(t_I^w-t) \right\rangle  =   \left\langle \Theta(t_M-t) \right\rangle    \left\langle  \Theta(t_I^w-t)  \right\rangle    = \Phi_I^w(t)\Phi_M(t) 
\eeq
since the indicator function $\Theta(t_M-t)\Theta(t_I^w-t) =1$ as long the infection persists, i.e. for $t<  t_{IM}^w =  {\rm min}(t_M,t_I^w)$ 
and we used independence of $t_I^w$ and $t_M$. (\ref{persist_infect}) has the interpretation of the persistence probability of the infection given the infection starts at $t=0$. 
Further, we will need the following quantities
\beq
\label{obs_rand}
\begin{array}{clr}
\ds {\cal R}(t) = \langle \Theta(t-t_I^w) \Theta(t_M-t_I^w) \rangle & = \ds \int_0^t b_r(\tau){\rm d}\tau = \int_0^t K_I^w(\tau)\Phi_M(\tau){\rm d}\tau & \\ \\
\ds {\cal D}(t) =  \langle \Theta(t-t_M) \Theta(t_I^w-t_M) \rangle &  = \ds \int_0^t b_d(\tau){\rm d}\tau   = \int_0^t K_M(\tau) \Phi_I^w(\tau){\rm d}\tau & \\ \\
\ds {\cal R}(t) + {\cal D}(t) = \int_0^{t} {\cal K}_{I,M}^w(\tau){\rm d}\tau  & = \ds \int_0^t[b_d(\tau)+b_r(\tau)]{\rm d}\tau  = 1 - \Phi_I^w(t)\Phi_M(t) & \\ \\
\ds {\cal D}(\infty) + {\cal R}(\infty) = 1 ,& 
\end{array}
\eeq
where is used $\Phi_M(0)=\Phi_I^w(0)=1$. The quantity ${\cal D}(\infty) \in (0,1)$ can be interpreted as the overall probability 
that a walker dies from its infection, and ${\cal R}(\infty)=1-{\cal D}(\infty) \in (0,1)$ as the probability that it survives the disease. 
For $t_M=\infty$ one has ${\cal D}(\infty)=0 $ and ${\cal R}(\infty)=1$ corresponding to zero mortality.
With these considerations, we can write the second equation of (\ref{evoleqsB}) as
\beq
\label{second_eq}
\frac{d}{dt}J_w(t) =\ds \frac{d}{dt} \left\{\int_0^t  {\cal A}_w(\tau) \Phi_M(t-\tau)\Phi_I^w(t-\tau){\rm d}\tau +J_w(0)\Phi_I^w(t)\Phi_M(t) \right\}. 
\eeq
For the numerical determination of the mean-field solutions, we used a fourth order Runge-Kutta (RK4) scheme in
the stochastic formulation of Eqs. (\ref{evoleqs}) together with the generalized Monte-Carlo convergence theorem
\beq
\label{MC_feature}
\lim_{n\to \infty} \frac{1}{n} \sum_{k=1}^n A(t-T^{(k)}) = 
\int_0^tA(t-\tau)K_i(\tau){\rm d}\tau 
\eeq
holding for random numbers $T^{(k)}$ drawn from PDF $K_i(\tau)$ for suitable causal functions $A$, avoiding in this way numerical evaluation of the 
involved convolutions.

\subsection{Zero mortality: endemic equilibrium}
\label{zero_mortality}
\begin{figure}[H]
\centerline{\includegraphics[width=0.65\textwidth]{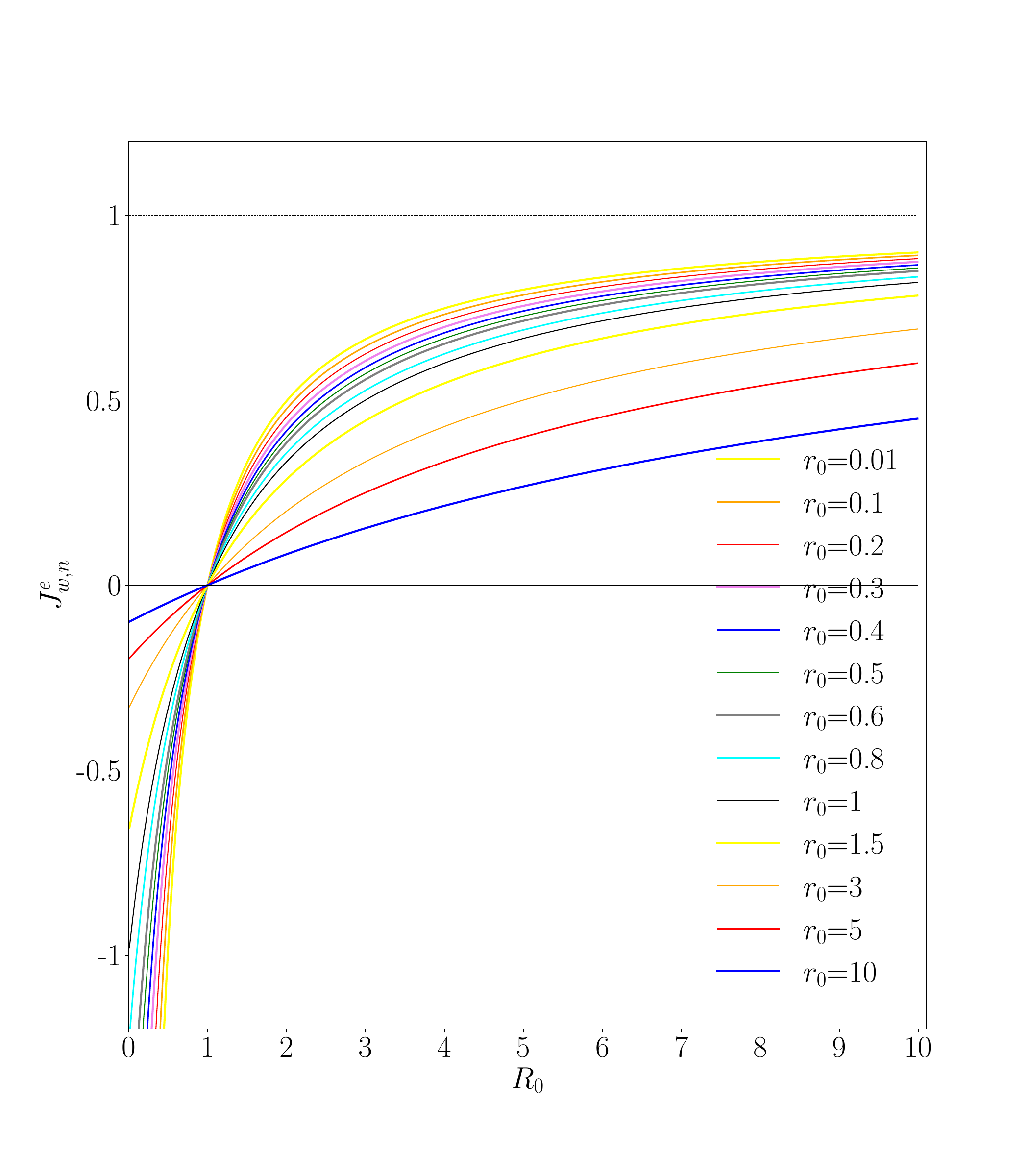}}
\caption{Endemic states $J_w^e, J_n^e = \frac{R_0-1}{R_0+r_0}$ versus $R_0$ of (\ref{endemic_values}) for some values of $r_0$. 
Read $r_0= \beta_n\langle t_I^n\rangle$ for $J_w^e$ and $r_0= \beta_w\langle t_I^w\rangle$ for $J_n^e$.}
\label{ENDEM_STATE}
\end{figure}

An important special case occurs for zero mortality $\frac{d}{dt} d_w(t)=0$ and is retrieved by $t_M=\infty$. Eqs. (\ref{evoleqs}) boil then down to
\beq
\label{evoleq-no-mort}
\begin{array}{clr}
\ds \frac{d}{dt} J_w(t) & = \ds  {\cal A}_w(t)    - \left\langle \,  {\cal A}_w(t-t_I^w)\, \right\rangle  
 -J_w(0)\left\langle \,\delta(t-t_I^w) \, \right\rangle     &   \\ \\
\ds \frac{d}{dt}J_n(t) & = \ds  {\cal A}_n(t)    - \left\langle \,  {\cal A}_w(t-t_I^n)\, \right\rangle  
 -J_n(0)\left\langle \,\delta(t-t_I^n) \, \right\rangle  ,  &
\end{array}
\eeq
where we skipped redundant equations.
In order to explore the existence of an endemic equilibrium which consists in constant asymptotic values $J_w(\infty), J_n(\infty)$, it is convenient to
consider the Laplace transformed system (\ref{evoleq-no-mort}) which reads
\beq
\label{evoleq-no-mort_LT}
\begin{array}{clr}
\ds \lambda{\hat J}_w(\lambda)  & = \ds  \left({\hat {\cal A}}_w(\lambda) +J_w(0)\right)\left[1-{\hat K}_I^w(\lambda)\right]    &   \\ \\
\ds \lambda{\hat J}_n(\lambda)   & = \ds   \left({\hat {\cal A}}_n(\lambda) +J_n(0)\right)\left[1-{\hat K}_I^n(\lambda)\right]  . &
\end{array}
\eeq
Here we have introduced the Laplace transform (LT) of a suitable function $f(t)$ as 
\beq
\label{LT_definition}
{\hat f}(\lambda) = \int_0^{\infty} e^{-\lambda t}f(t){\rm d}t .
\eeq
The endemic equilibrium solves (\ref{evoleq-no-mort}) for vanishing left-hand sides and representing the asymptotic solution for $t \to \infty$.
Letting $\lambda \to 0$ in (\ref{evoleq-no-mort_LT}), by using the limiting value theorem 
\beq
\label{lim_theo}
f(\infty) = \lim_{\lambda \to 0}\lambda {\hat f}(\lambda)
\eeq
we can conveniently extract the endemic values
\beq
\label{evoleq-no-mort_endemic}
\begin{array}{clr}
\ds J_w^e  & = \ds   {\cal A}_w(\infty) \langle t_I^w\rangle =  \beta_w \langle t_I^w\rangle    (1-J_w^e)J_n^e  &   \\ \\
\ds J_n^e   & = \ds    {\cal A}_n(\infty) \langle t_I^n\rangle  = \beta_n  \langle t_I^n\rangle  (1-J_n^e)J_w^e     &
\end{array}
\hspace{1cm} {\cal A}_{w,n}(\infty) = \beta_{w,n} S_{w,n}^e J_{n,w}^e\, , 
\eeq
independent of initial conditions $J_{w,n}(0)$, where we denote the endemic values from now on $ J_{w,n}^e=J_{w,n}(\infty)$, $S_{w,n}^e=S_{w,n}(\infty)= 1-J_{w,n}^e$.
In these relations, we assume that the PDFs $K_{I}^{w,n}(t)$ are such that $\langle t_I^{w,n} \rangle < \infty$. One gets straightforwardly
\beq
\label{endemic_values}
\begin{array}{clr}
\ds J_w^e  & = \ds  \frac{R_0-1}{R_0} \frac{\beta_w \langle t_I^w\rangle}{1+ \beta_w \langle t_I^w\rangle} & \\ \\
\ds J_n^e   & = \ds  \frac{R_0-1}{R_0} \frac{\beta_n \langle t_I^n\rangle}{1+ \beta_n \langle t_I^n\rangle}  &  
\end{array} \hspace{1cm} R_0 = \beta_w \langle t_I^w \rangle  \beta_n \langle t_I^n \rangle = \frac{1}{S_w^eS_n^e} .
\eeq
These relations have exchange symmetry with respect to walkers and nodes, reflecting this symmetry in
the system (\ref{evoleq-no-mort}).
Since $J_{w,n}^e \in (0,1)$ the endemic equilibrium exists only for $R_0 >1$.
Worthy of mention are the limits
\\[2ex]
(i) $\beta_w\langle t_I^w \rangle \to \infty$ (while $\beta_n\langle t_I^n\rangle $ finite): 
$J_w^e \to 1$, whereas $J_n^e \to 
\frac{\beta_n\langle t_I^n \rangle}{1+\beta_n\langle t_I^n \rangle} < 1$
\\[2ex] (ii) $\beta_n\langle t_I^n \rangle \to \infty$ (while $\beta_w\langle t_I^w\rangle $ finite): $J_n^e \to 1$ whereas $J_w^e \to 
\frac{\beta_w\langle t_I^w \rangle}{1+\beta_w\langle t_I^w \rangle} < 1 , $
\\[2ex]
where in both cases $R_0 \to \infty$. 
We depict the $R_0$-dependence of the endemic equilibrium in Fig. \ref{ENDEM_STATE}. 
Only positive values of $J_w^e, J_n^e$ occurring for $R_0>1$ correspond to endemic equilibria.
We infer that the healthy state $S_{w,n}=1$ is unstable for $R_0>1$ and the endemic state (\ref{endemic_values}) stable. On the other hand, for $R_0<1$ the healthy state is globally stable, and the endemic state does not exist.
$R_0 =  \beta_w \langle t_I^w \rangle  \beta_n \langle t_I^n \rangle >1$ is hence the condition that the disease without mortality is starting to spread. 
This can be proven by a stability analysis of endemic and healthy states, see \cite{SISI_Entropy_2024} for details, and also follows from the results of the next section when considering the limit of zero mortality.

\subsection{Spreading condition with mortality}
\begin{figure}[H]
\centerline{\includegraphics[width=0.65\textwidth]{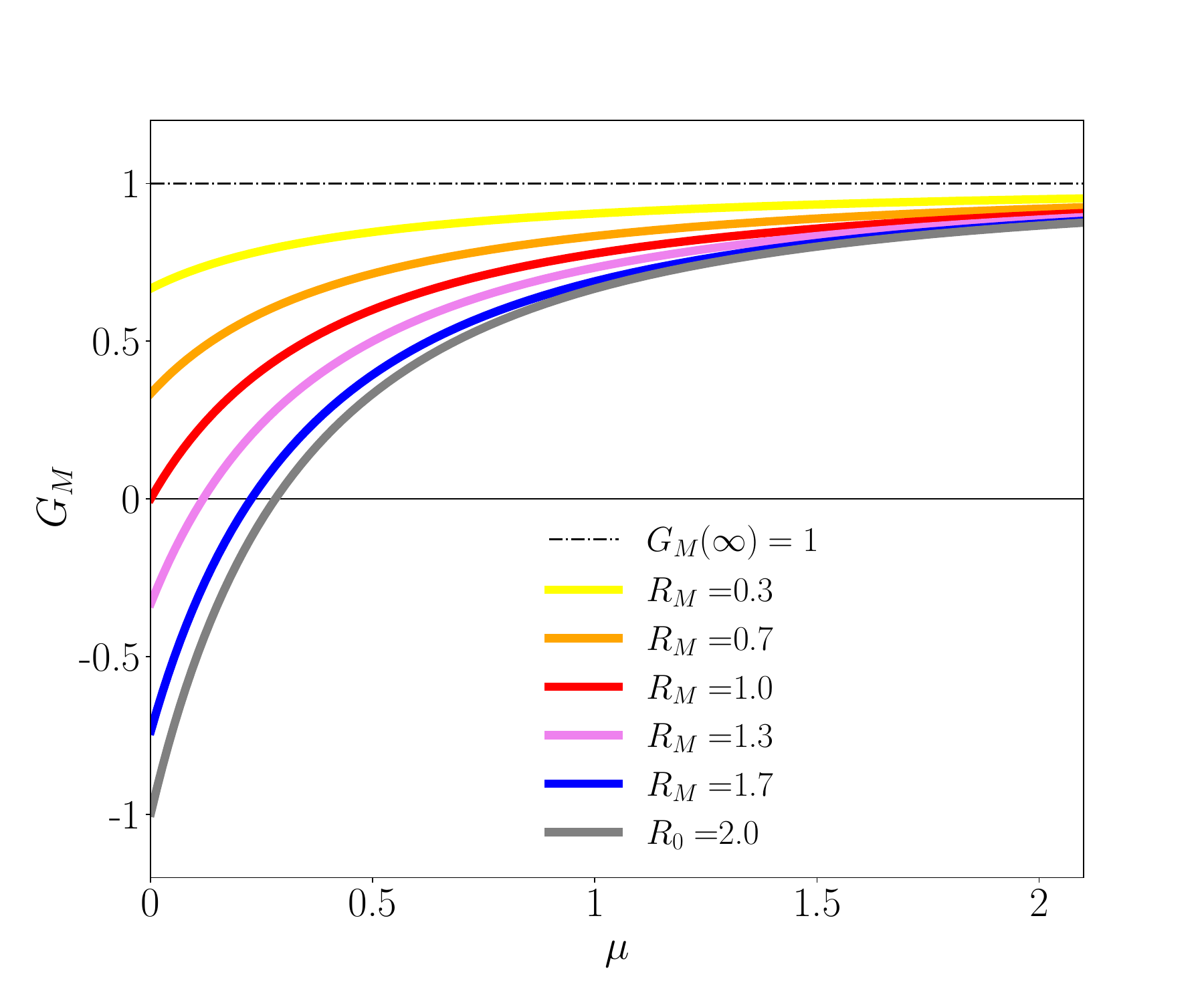}}
\caption{We depict $G_M(\mu)$ from Eq. (\ref{G-M-function}) for some $R_M$
for exponentially distributed $t_I^{w,n}, t_M$ ($\Phi_I^{w,n}(t)= e^{-\xi_I^{w,n}t}$, $\Phi_M(t)=e^{-\xi_M t}$).
The parameters are $\beta_w=0.5$, $\beta_n=2$, $\xi_I^w=1$, $\xi_I^n=0.5$. The basic reproduction number without mortality ($\xi_M=0$) is $R_0=2$ 
with $R_M = R_0/(1+\xi_M)$. One can see that (\ref{G-M-function}) has a positive solution for $R_M>1$ indicating instability of the healthy state.}
\label{GM_plot}
\end{figure}

Here we are interested in how mortality modifies the basic reproduction number. From now on we denote with $R_0$ the basic reproduction number without mortality (see (\ref{endemic_values}) and with $R_M$ the basic reproduction number in presence of mortality. To answer this question, we perform a linear stability analysis of the healthy state $S_{w,n}=1$.
To that end we set
\beq
\label{lin-stab-mort}
S_w(t) =1 + a \, e^{\mu t}, \hspace{0.5cm} J_w(t)= b \, e^{\mu t} , \hspace{0.5cm} d_w(t) = -(a+b)\, e^{\mu t} , 
\hspace{0.5cm} S_n(t) = 1 - c e^{\mu t}, \hspace{0.5cm} J_n(t) =  c e^{\mu t}
\eeq
with ${\cal A}_w(t) = \beta_w c\, e^{\mu t}$ and ${\cal A}_n(t) = \beta_n b\, e^{\mu t}$, where $a,b,c,$ are 'small' constants. 
Note that (\ref{lin-stab-mort}) is such that $S_w(t)+J_w(t)+d_w(t)=1$ and $S_n(t)+J_n(t)=1$.
Plugging this expression for $\mu \geq 0$ into three independent Eqs. of (\ref{evoleqs}), one gets
\beq
\label{mat_equations_mort}
\left[\begin{array}{lll} \mu \, ;& \hspace{0.5cm} 0 ;& \beta_w[1-{\hat b}_r(\mu)] ]   \\ \\
\mu ;& \hspace{0.5cm} \mu \, ;& \beta_w {\hat b}_d(\mu) \\ \\
 0 \, &  -\beta_n[1-{\bar K}_I^n(\mu)] \,  ;  & \mu
 \end{array}\right] \cdot \left(\begin{array}{l} a \\ \\ b \\ \\ c \end{array}\right) = \left(\begin{array}{l} 0 \\ \\ 0 \\ \\0 \end{array}\right) .
\eeq
Putting the determinant of this matrix to zero leads to 
\beq
\label{G-M-function}
G_M(\mu) = 1- \beta_n\beta_w\frac{[1-{\bar K}_I^n(\mu)]}{\mu}\frac{[1-{\hat {\mathcal K}}^w_{I,M}(\mu)]}{\mu} = 0 ,
\eeq
where $ {\cal K}_{I,M}^w(t) = b_d(t)+b_r(t) = -\frac{d}{dt}[\Phi_I^w(t)\Phi_M(t)]$ is the PDF that governs the end of the infection (either by
recovery or by death). In particular,
\beq
\label{in_par}
\frac{[1-{\hat {\cal K}}^w_{I,M}(\mu)]}{\mu} = \int_0^{\infty}\Phi_M(t)\Phi_I^w(t) e^{-\mu t}{\rm d}t 
\eeq
is the LT of the persistence probability $\Phi_M(t)\Phi_I^w(t) =\langle \Theta(t_M-t) \Theta(t_I^w-t)\rangle $ of the walker's infection, i.e. the probability that $t< {\rm min}(t_I^w,t_M)$. The healthy state is unstable if there is a solution $\mu>0$ solving (\ref{G-M-function}).
In the following 
the mean sojourn time of a walker in compartment I with mortality 
\beq
\label{ROM_res}
\langle t^w_{IM} \rangle  = \langle  {\rm min}(t_I^w,t_M) \rangle  =  \frac{[1-{\hat {\mathcal K}}^w_{I,M}(\mu)]}{\mu}\bigg|_{\mu=0} =\int_0^{\infty} t  {\cal K}_{I,M}^w(t){\rm d}t =
\int_0^{\infty}  \Phi_M(t)\Phi_I^w(t) {\rm d}t \, \leq \,\int_0^{\infty}  \Phi_I^w(t) {\rm d}t = \langle t_I^w \rangle
\eeq
comes into play, leading to
\beq
\label{G_M_mu_zero}
G_M(\mu)\bigg|_{\mu=0} = 1- \beta_n\beta_w \langle t_I^n \rangle \langle t^w_{IM} \rangle  .
\eeq
From (\ref{ROM_res}) follows that 
$\langle t^w_{IM} \rangle  \leq \langle t_I^w \rangle $ (equality only for zero mortality) reflecting simply the feature that 
$t_{IM}^w = {\rm min}(t_I^w,t_M) \leq t_I^w$ with again equality for zero mortality $t_M=\infty$.
On the other hand, we have $G_M(\mu) \to 1$ for $\mu \to \infty$, so there is a positive solution $\mu=\mu_*$ of $G_M(\mu)=0$ only if 
\beq
\label{basic_rep_with_mortality}
R_M =\beta_n\beta_w \langle t_I^n \rangle \langle t^w_{IM} \rangle \, > \, 1, 
\eeq
see Fig. \ref{GM_plot}. We identify the quantity $R_M$ as the basic reproduction number modified by mortality, which fulfills
\beq
\label{R_M_R_0}
R_M  = \beta_n\beta_w \langle t_I^n \rangle \langle t^w_{IM} \rangle \leq   R_0 = \beta_n\beta_w \langle t_I^n 
\rangle \langle t_I^w \rangle
\eeq
(equality for zero mortality only).
%
%\begin{figure}[b]
%
To visualize the effect of mortality on the instability of the healthy state, we plot $G_M(\mu)$ for a few values
of $R_M$ in Fig. \ref{GM_plot}. One can see that increasing the mortality rate parameter $\xi_M$ sufficiently turns an unstable healthy state into a stable one. $R_M$ is monotonously decreasing 
with increasing $\xi_M$ assuming here an exponentially distributed $t_M$ following the PDF $K_M(t) =\xi_M e^{-\xi_M t}$.

\section{Random Walk Simulations}
\label{RW_simulations}

\subsection{Multiple walkers approach}
\label{multiple_walkers}

In order to obtain a microscopic picture of the dynamics, we implement the mean field model into random walk simulations. 
We consider a population of $Z$ individuals, represented by random walkers moving independently on connected undirected (ergodic) networks\footnote{Each pair of distinct nodes is connected by a path along edges of finite length.}, where we focus on random graphs of the Barabási–Albert (BA) and Watts–
Strogatz (WS) type by using the PYTHON NetworkX  library.

We explore the
effect of the graph topology on the propagation of the disease and compare the results of the random walk simulations with those of
the mean field model by integrating evolution Eqs. (\ref{evoleqs}) numerically in a fourth-order
Runge–Kutta scheme (RK4). Doing so, we employ feature (\ref{MC_feature}) to avoid performing the convolutions explicitly. At the end of the section, we also investigate how confinement measures may affect the spreading of the disease.

\begin{figure}[H]
\centerline{
\includegraphics[width=0.3\textwidth]{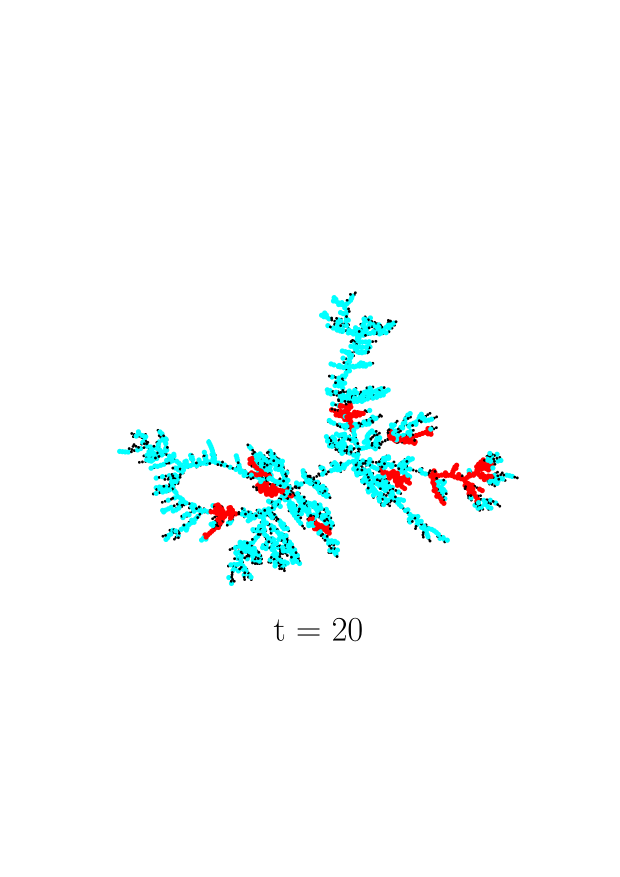}
\includegraphics[width=0.3\textwidth]{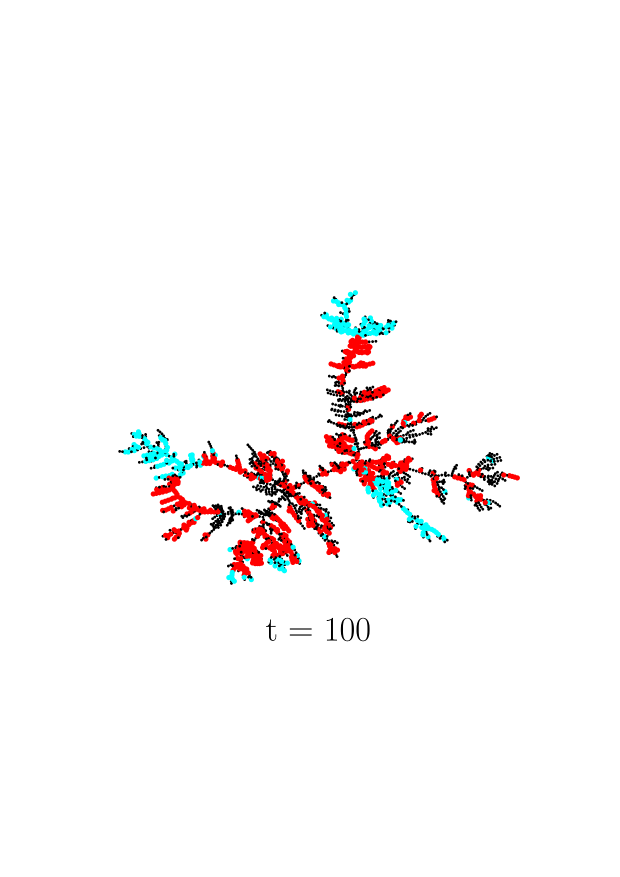}
}
\vspace{-10mm}
\centerline{
\includegraphics[width=0.3\textwidth]{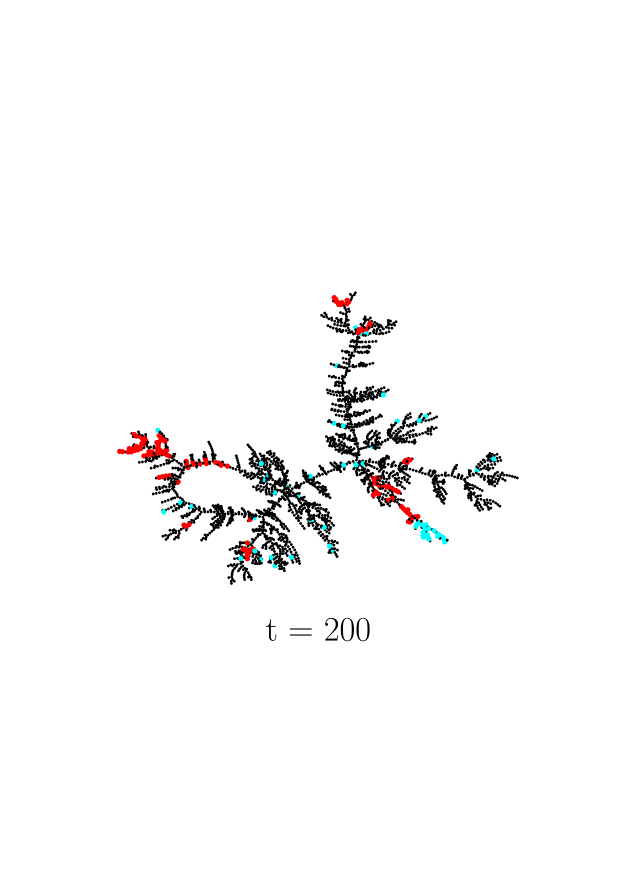}
\includegraphics[width=0.3\textwidth]{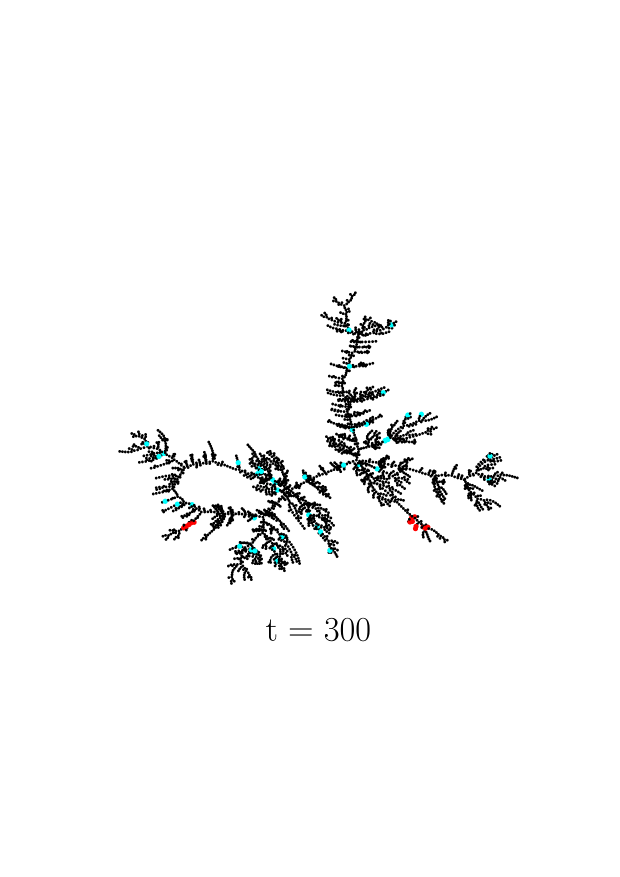}
}
\vspace{-10mm}
\centerline{
\includegraphics[width=0.5\textwidth]{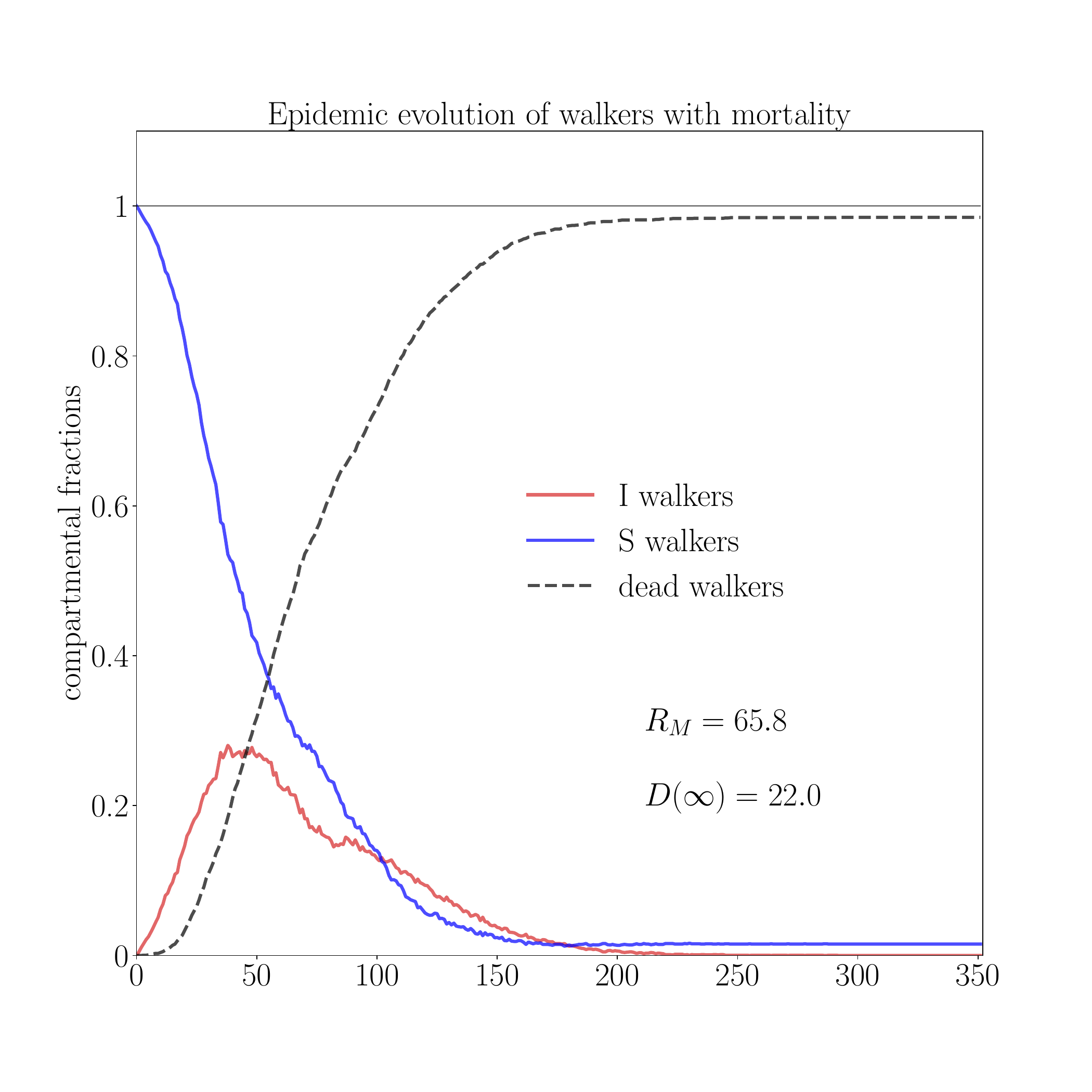}
\includegraphics[width=0.5\textwidth]{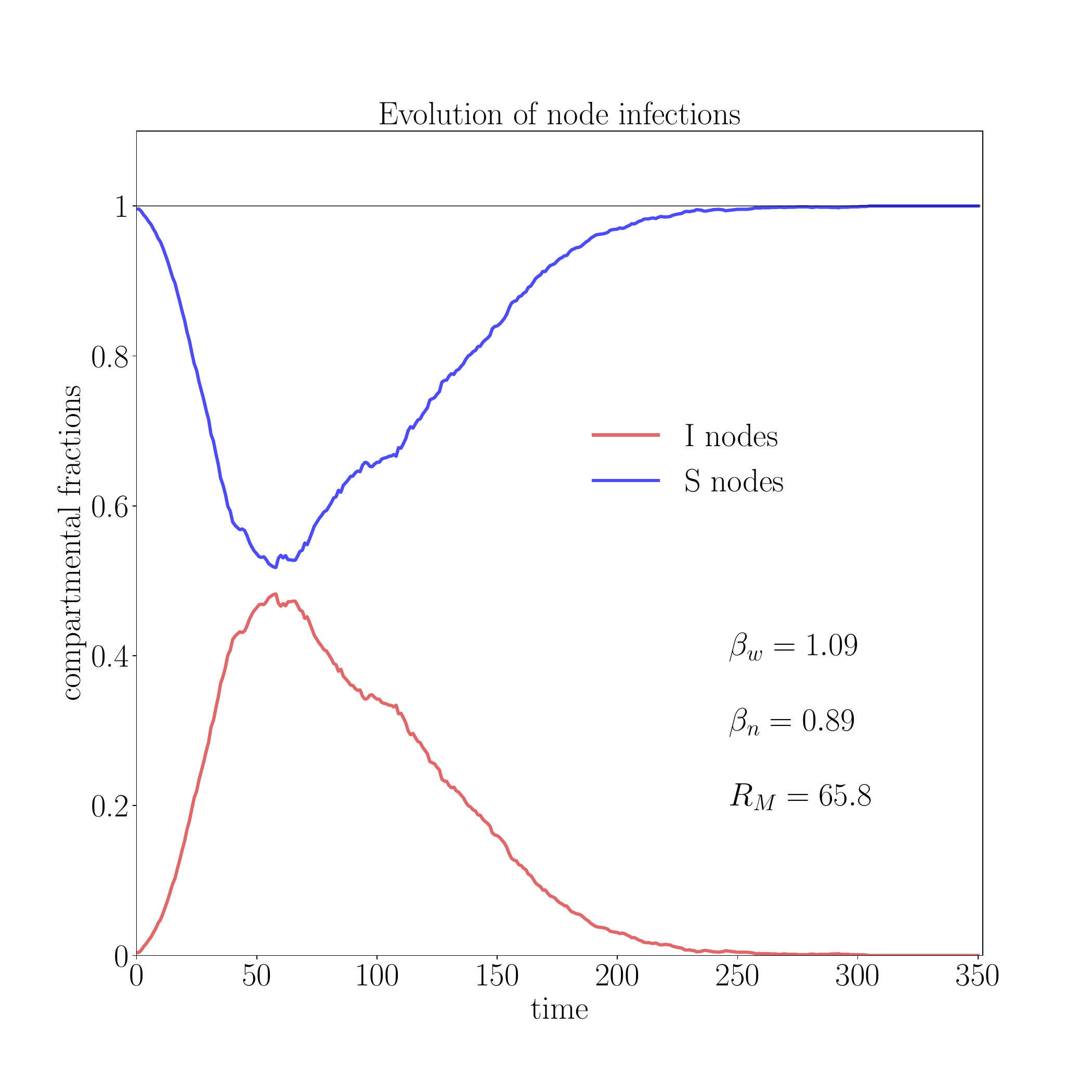}
}
\caption{Spreading with mortality for a single random walk realization (for PYTHON $seed=0$) in the WS graph 
$G= nx.connected\_watts\_strogatz\_graph(2500, m=2, p=0.8, seed=seed)$ ($Z=2500$ walkers, $N=2500$ nodes; average degree $\langle k \rangle = \frac{1}{N}\sum_{i=1}^N k_i = 2=m$). The parameters
of the Gamma distributed compartment sojourn times are:
  $\langle  t_I^n\rangle =12$, $\xi_I^n=10^5$ ($t_I^n$ sharp $\delta$-distributed); 
  $\langle t_I^w \rangle =6$, $\xi_I^w=10$, $\langle t_M\rangle =10$, $\xi_M=0.4$. Initial condition: $10$ infected nodes at $t=0$. An animated simulation video of this dynamics can be 
  \href{https://drive.google.com/file/d/16zAzVnUQZwLu8PdrJBbY71AngIT33iRQ/view?usp=sharing}{\textcolor{blue}{viewed by clicking online here}}. S walkers are represented in cyan color, I walkers in red. Nodes are drawn in black without representation of their infection state. }
\label{WS_spreading}
\end{figure}

\begin{figure}[H]
\centerline{
\includegraphics[width=0.55\textwidth]{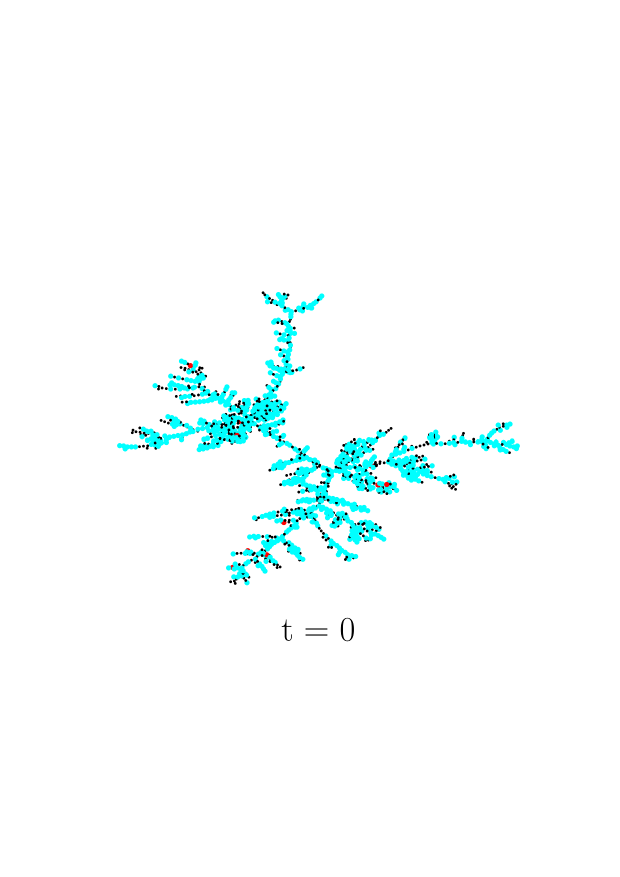}
}
\vspace{-10mm}
\centerline{
\includegraphics[width=0.55\textwidth]{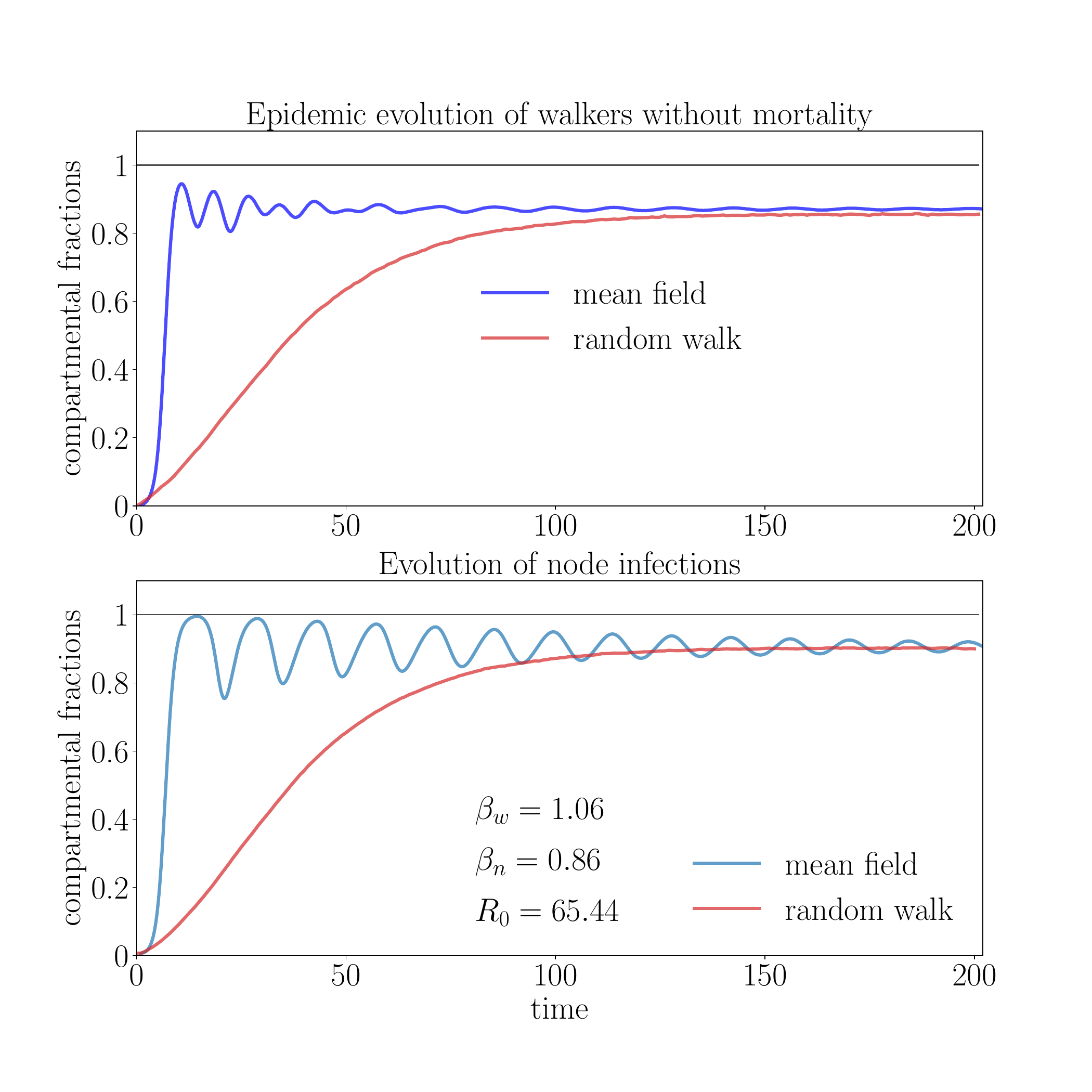}
\includegraphics[width=0.55\textwidth]{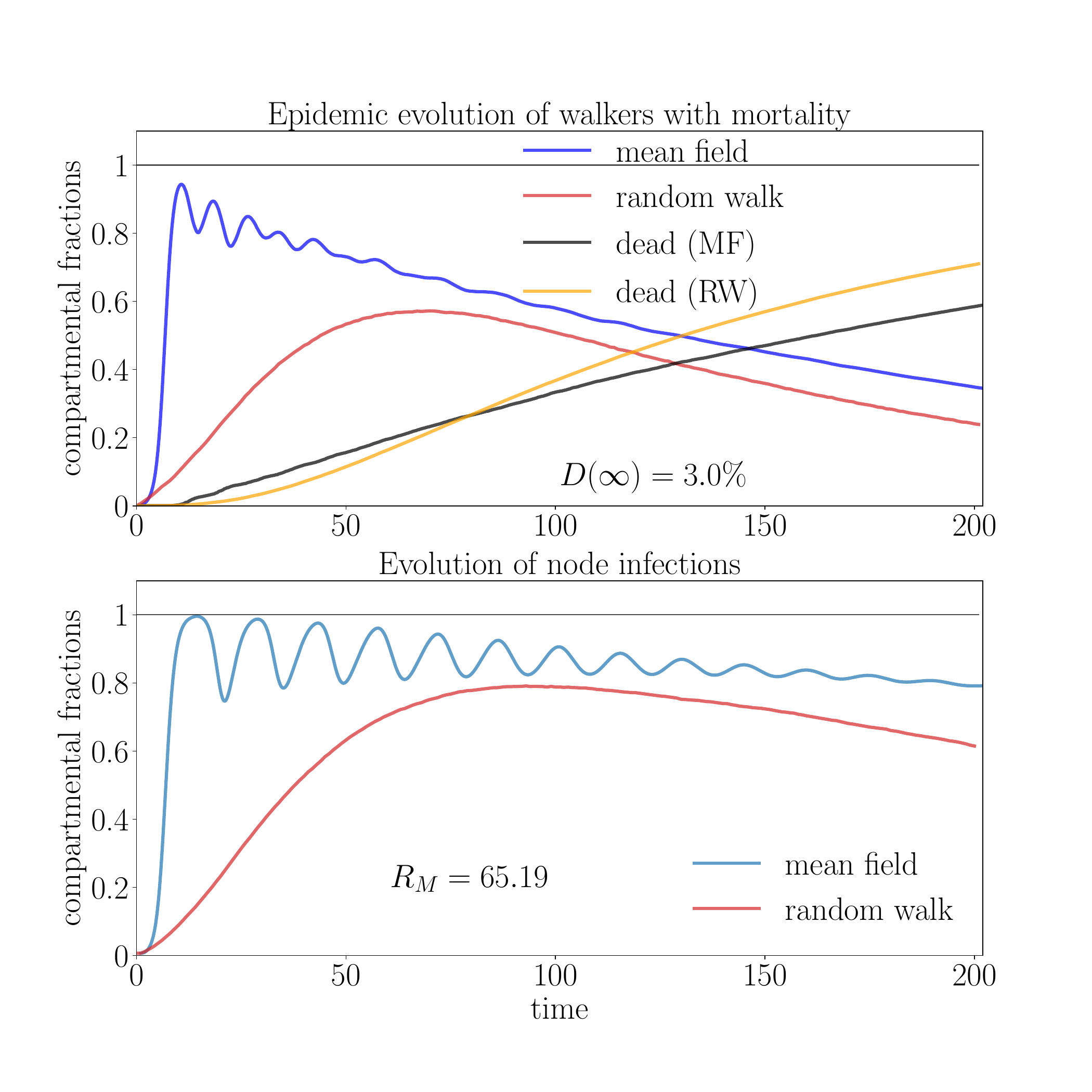}
}
\vspace{-5mm}
\caption{Spreading in the WS graph [$nx.connected\_watts\_strogatz\_graph(1500, m=2, p= 0.9, seed=seed)$] 
with a higher rewiring probability $p=0.9$ than in Fig. \ref{WS_spreading} and with $1500$ nodes and $Z=1500$ walkers ($10$ I nodes at $t=0$.), average degree $\langle k\rangle =m = 2$. Left frame: without mortality. Right frame: same setting with mortality. Average over $100$ equivalent random walk realizations (by choosing for each realization a different PYTHON seed $= \{ 0,\ldots, 99 \}$). All other parameters are identical as in Fig. \ref{WS_spreading}  except here we have $\xi_M= 2$. The simulations confirm that $ R_M = 65.19 < R_0 = 65.44$. \label{WS_mort_nomort}}
\end{figure}

The considered WS network has connectivity parameter $m=2$ with emergence of a large-world graph, see Figs. \ref{WS_spreading}, \ref{WS_mort_nomort}. 
The WS graph is generated by starting in a first step by a ring of $N$ nodes, where each node is connected to its $m$ left and $m$ right closest neighbors, thus each node has $2m$ connections. In a second step, each of the connections is rewired with a probability $p$ to a randomly chosen node by avoiding multiple connections \cite{ErdoesRen1960,WattsStrogats1998}.
The considered WS graph $G$ is generated by NetworkX employing $G = nx.connected\_watts\_strogatz\_graph(N, m, p, seed=seed)$ (allowing to fix the "seed", i.e. the generated random numbers to obtain the same random graph in different simulation runs).

\begin{figure}[H]
\centerline{
\includegraphics[width=0.55\textwidth]{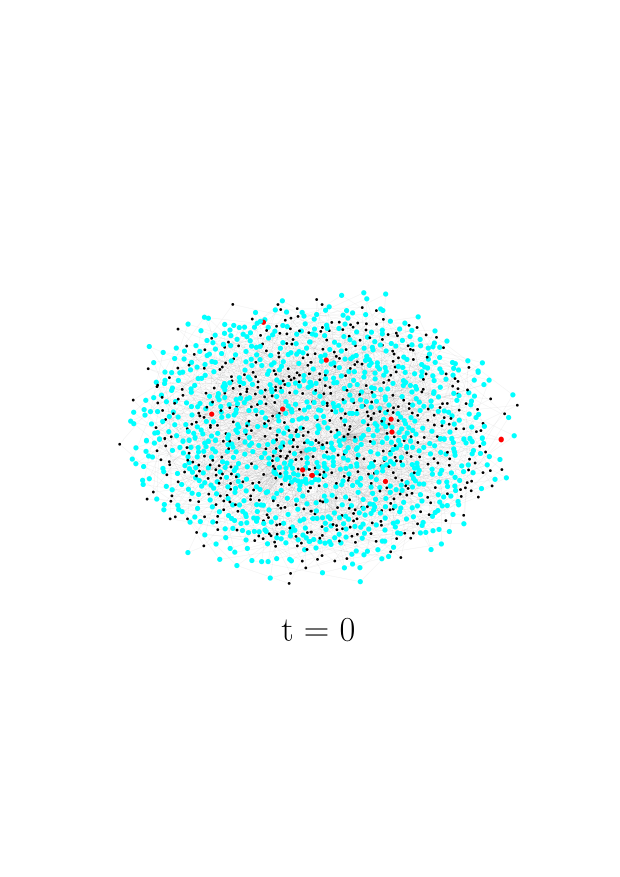}
}
\vspace{-10mm}
\centerline{
\includegraphics[width=0.55\textwidth]{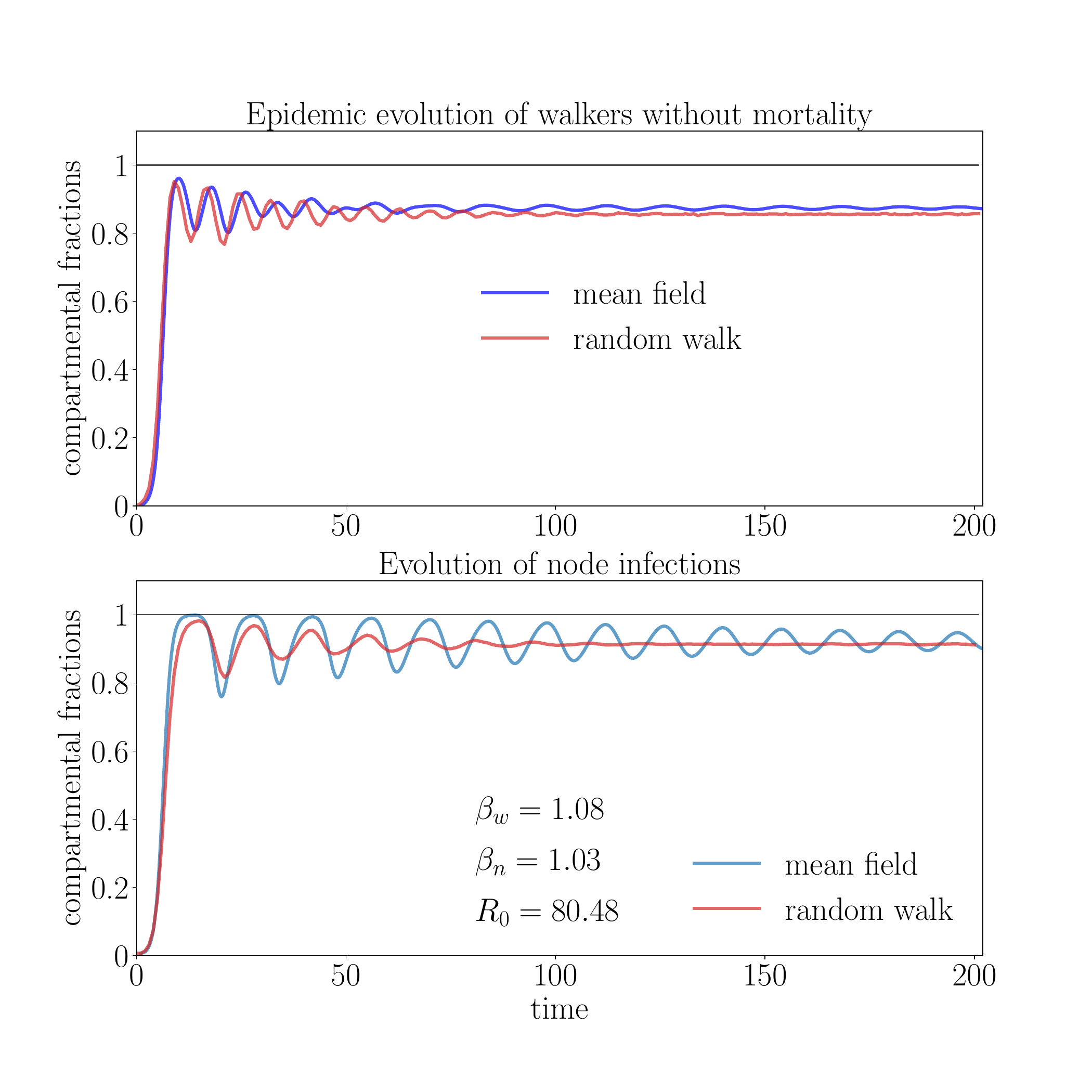}
\includegraphics[width=0.55\textwidth]{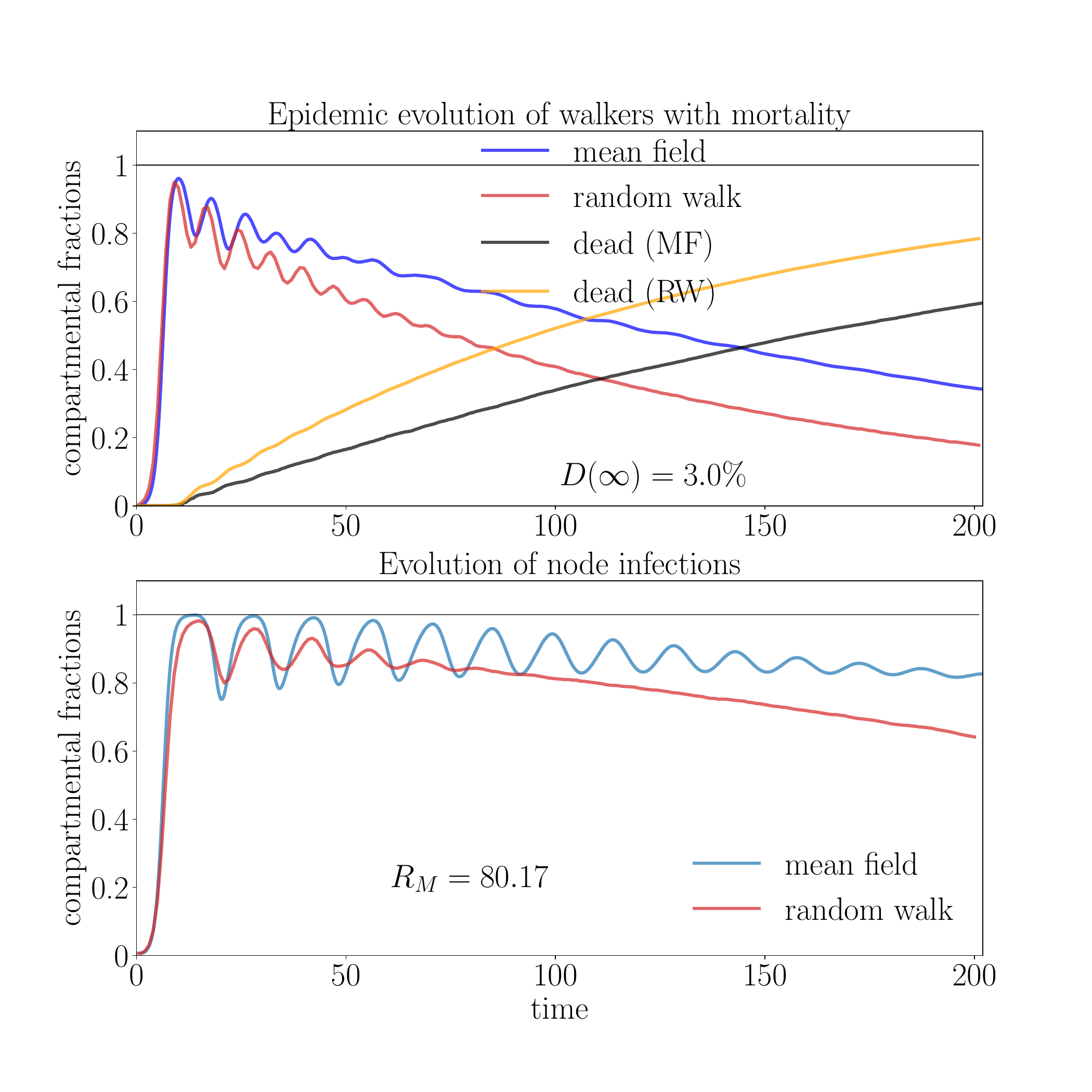}
}
\caption{Spreading in the BA graph [$nx.barabasi\_albert\_graph(1500, m=6, seed=seed)$]
 ($10$ I nodes at $t=0$) with $1500$ walkers, average degree $\langle k\rangle = 11.95$. Left frame: without mortality. Right frame: same setting with mortality. Average over $100$ equivalent random walk realizations. The Gamma-distributed waiting time parameters are identical as in Fig. \ref{WS_mort_nomort}. Basic reproduction numbers with and without mortality: $ R_M \approx 80.17 < R_0 \approx 80.48 $. }
\label{BA_mort_nomort}
\end{figure}

To explore the effect of the connectivity of a network, we consider the spreading in a BA network (generated by $G = nx.barabasi\_albert\_graph(N, m, seed=seed)$)
which is small world (Fig. \ref{BA_mort_nomort}) contrarily to the considered WS graph. The BA graph is generated by a preferential attachment mechanism for newly added
nodes: One starts with $m_0$ connected nodes and adds new nodes in the following way. Any newly added node
is connected with $m \leq m_0$ existing nodes ($m$ is called attachment parameter). The newly added node is preferentially connected to a node with high degree.
In this way, an asymptotically scale-free small-world graph emerges with a power law degree distribution \cite{Barabasi2016,BarabasiAlbert1999,Jeong-etal200,BarabasiAlbertJeong1999}.
It has been pointed out in the literature that small-world architectures are favorable
for the propagation of epidemics \cite{Satoras-Vespignani-etal2015,Pastor-SatorrasVespignani2001}. This is also confirmed by the case study of this section.

We employ Gamma distributed $t_I^{w,n},t_M$ waiting-times due to the high flexibility of the Gamma distribution to adapt relevant shapes observed in real world situations.
The Gamma PDF reads
\beq
\label{Gamm_dist}
K_{\alpha,\xi}(t) = \frac{\xi^{\alpha}t^{\alpha-1}}{\Gamma(\alpha)} e^{-\xi t}, \hspace{1cm}\xi ,\,\,  \alpha > 0 ,
\eeq
where $\alpha$ is the so-called ``shape parameter'' and $\xi$ the rate parameter (often, the term ``scale parameter'' is used, $\theta=\xi^{-1}$) and $\Gamma(\alpha)$ stands for the Gamma function. The Gamma PDF has finite mean $\langle t \rangle_{\alpha,\xi} = \int_0^{\infty} t K_{\alpha,\xi}(t){\rm d}t = \frac{\alpha}{\xi}$, and
for $\alpha \in (0,1)$, the Gamma PDF is weakly singular at $t=0$ and $\alpha=1$ recovers exponential PDFs. For $\alpha >1$, the Gamma PDF has a maximum at $t_{max}=\frac{\alpha-1}{\xi}$, and becomes narrower the larger $\xi$
approaching a Dirac $\delta$-distribution $ K_{\alpha,\xi}(t) \to \delta(t-t_0)$ for $\xi \to \infty$ while keeping its mean $\langle t \rangle_{\alpha,\xi}= t_0$ constant.

In order to describe the random walk of each walker, 
we denote with $i=1,\ldots N$ the nodes of the network and introduce the $N\times N$ the symmetric adjacency matrix $(A_{ij})$, where $A_{ij}=1$ if the pair of nodes $i,j$ is connected by an edge, and $A_{ij}=0$ if the pair is disconnected. 
Further, we assume $A_{ii}=0$ to avoid self-connections of nodes. We restrict our analysis to undirected networks, where edges have no predefined direction and the adjacency matrix is symmetric. The degree $k_i$ of a node $i$ counts the number
of its neighbor nodes (connected with $i$ by edges).
Each walker performs independent Markovian steps between connected nodes. 
 The steps from a node $i$ to one of its $k_i=\sum_{j=1}^NA_{ij}$ neighbor nodes are chosen with probability $1/k_i$, 
 leading for all walkers to the same transition matrix which reads \cite{fractional_book_MiRia2019,Newman2010,NohRieger2004}
\beq
\label{markovian walk}
\Pi(i \to j) = \frac{A_{ij}}{k_i} , \hspace{1cm} z=1,\ldots, Z , \hspace{1cm} i,j=1,\ldots, N
\eeq
and is by construction row-normalized $\sum_{j=1}^N\Pi(i \to j) =1$. To ensure ergodicity, we exclude that the graph is bipartite by
the presence of at least one return path of odd length \cite{NohRieger2004}.
In the simulations, the departure nodes at $t=0$ of each walker is randomly chosen. 
The path of each walker is independent and not affected by contacts
or infection events by other walkers. We assume uniform transmission probabilities $p_{w,n}=1$ of walkers and nodes unless stated otherwise.

\begin{figure}[H]
\centerline{
\includegraphics[width=0.55\textwidth]{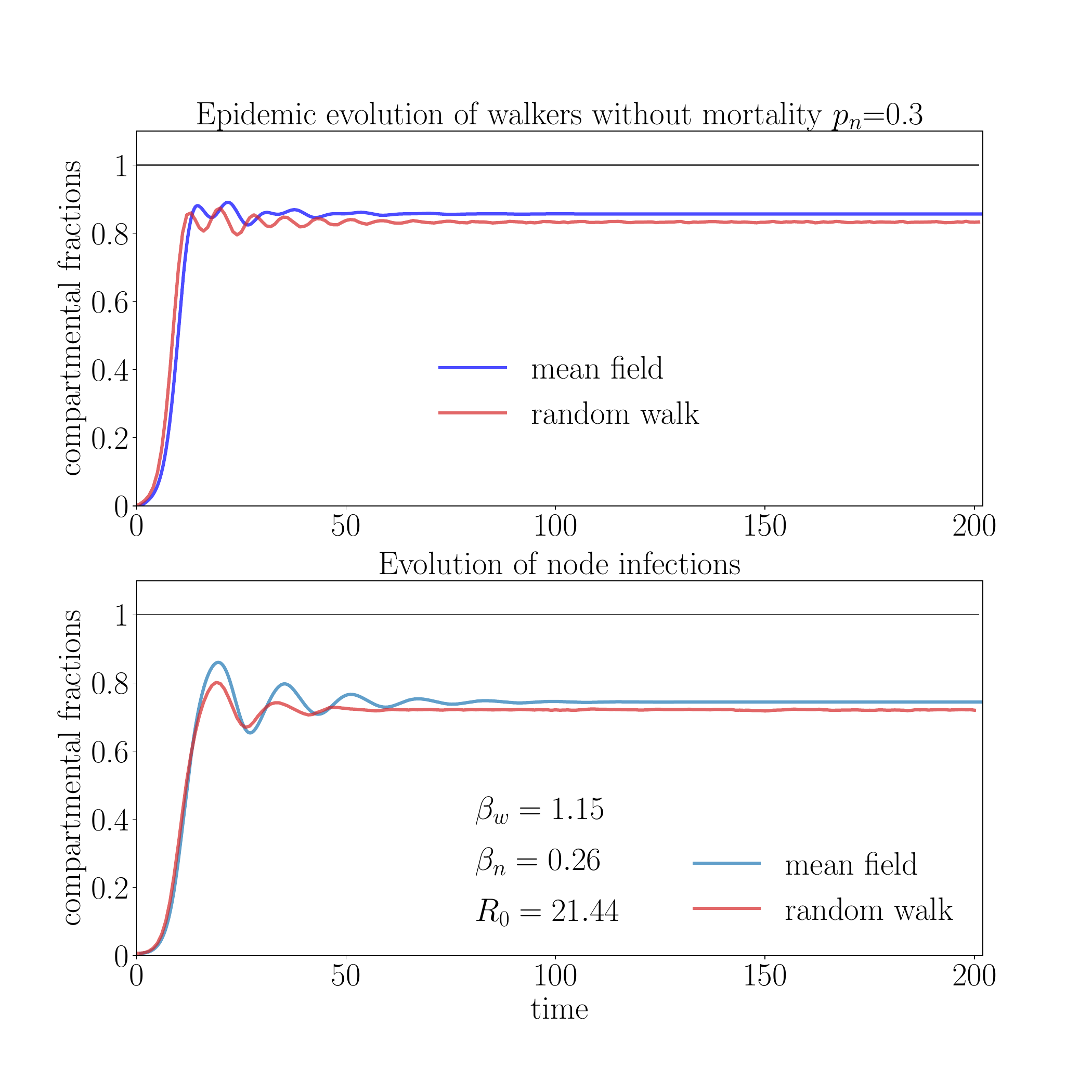}
\includegraphics[width=0.55\textwidth]{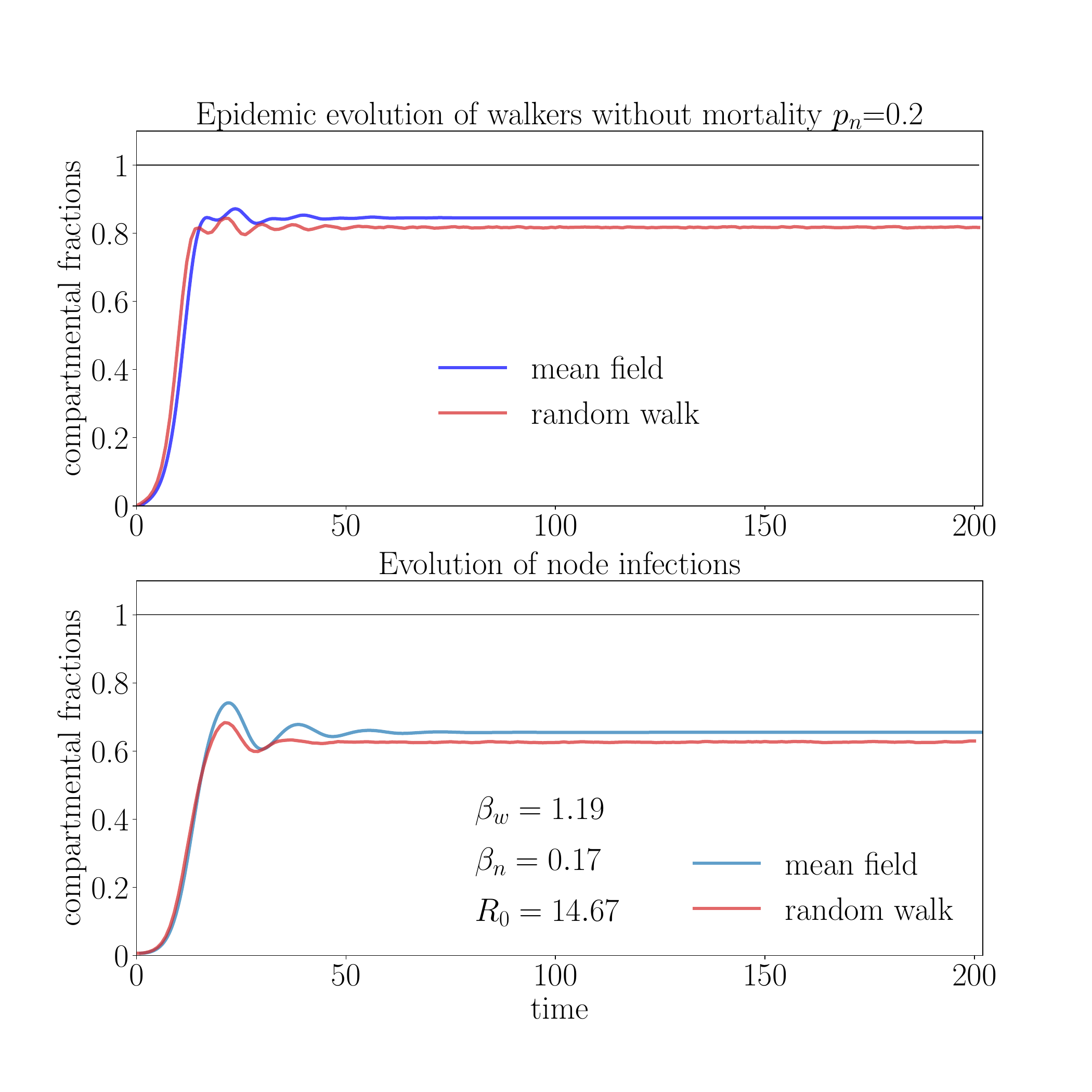}
}
\centerline{
\includegraphics[width=0.55\textwidth]{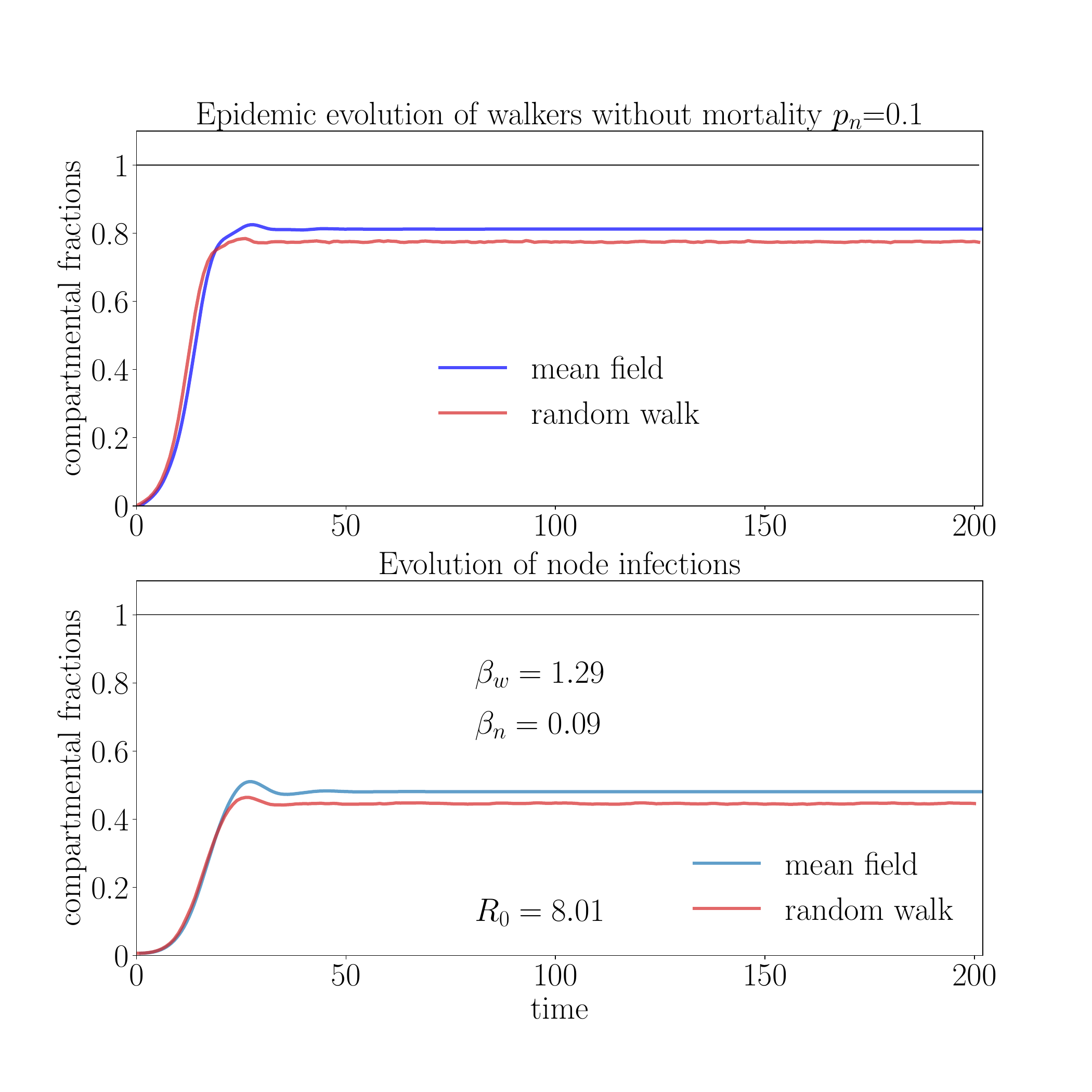}
\includegraphics[width=0.55\textwidth]{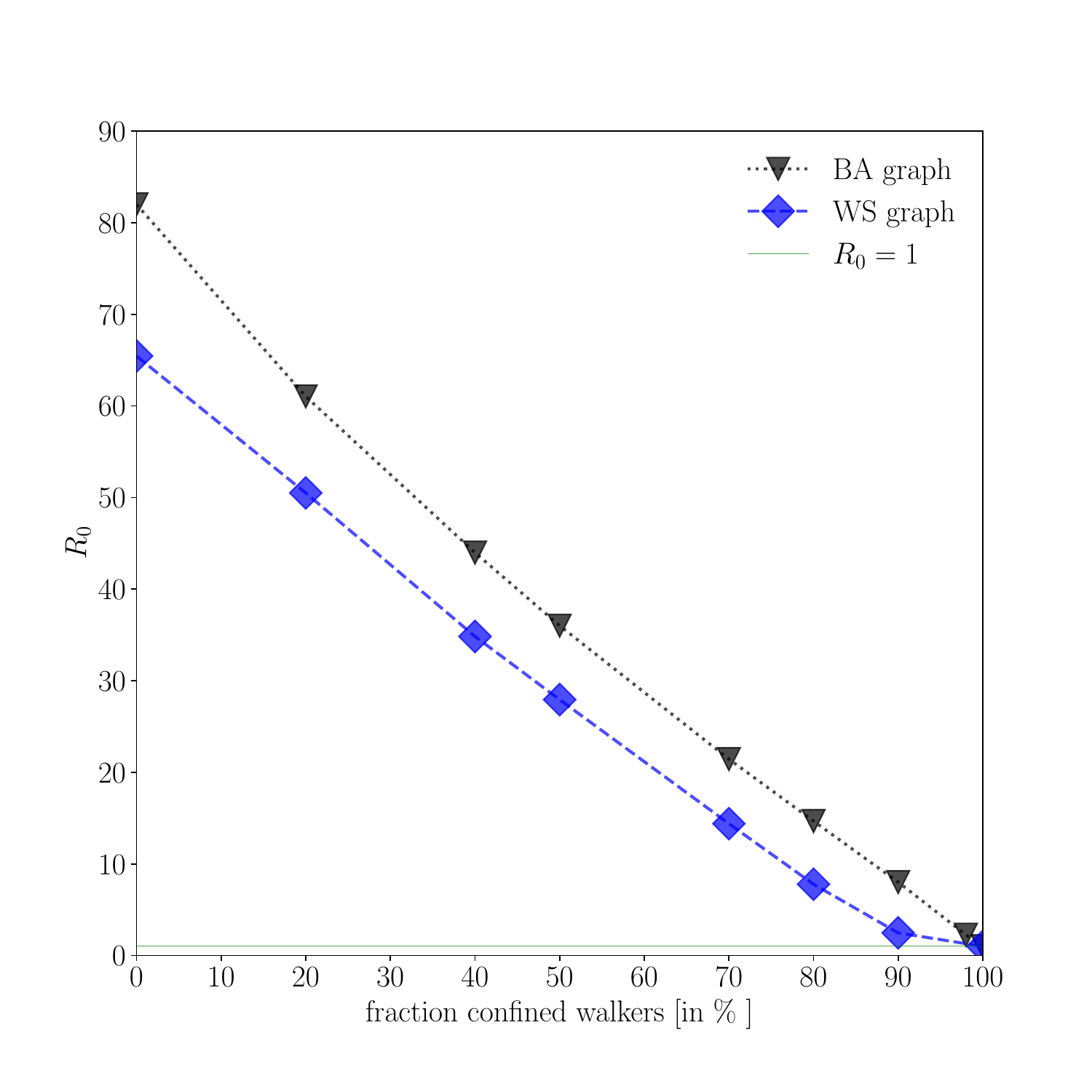}
}
\caption{Spreading without mortality in the same BA graph as in Fig. \ref{BA_mort_nomort} ($1500$ walkers and averaging over $100$ equivalent RW realizations) 
and the same setting with identical parameters of Gamma distributed $t_I^{w,n}$ for some values of transmission probabilities $p_n$ of I-walkers to S-nodes (lower left and upper frames). 
This mimics the effect of confinement measures, where $1-p_n$ is identified with the fraction of confined walkers' population. \\
Lower right frame: $R_0$ versus fraction of confined walkers for the BA graph of this figure and for the WS graph of Fig. \ref{WS_mort_nomort}.}
\label{BA_mort_nomort_conf}
\end{figure}

\subsection{Discussion}
\label{discussion}
Fig. \ref{WS_spreading} shows the epidemic evolution on a large world WS graph with high mortality.
The overall probability of $D(\infty) \approx 0.22$ that an infected walker dies in one infection leads to an extremely high
mortality in the walkers' population due to successive multiple infections of surviving walkers. One can see in the animated video 
\href{https://drive.google.com/file/d/16zAzVnUQZwLu8PdrJBbY71AngIT33iRQ/view?usp=sharing}{\textcolor{blue}{
 -- to view it click online here}} -- that almost the entire walkers' population is dying during the epidemic. 
 A vector transmitted disease with such a high mortality is for instance Pestilence.

We observe that the mean field model agrees well with the random walk result in the case of the small world BA graph (Figs. \ref{BA_mort_nomort},  
\ref{BA_mort_nomort_conf}). For the WS graph of Fig. \ref{WS_mort_nomort} which is large world, the agreement between random walk and mean field is characterized by a delayed increase of the compartment fractions in the random walk simulations which can be explained that the infection needs more time to spread in a weakly connected architecture with long average distances between nodes.
For large observation times, the agreement is well
for zero mortality, but less well in presence of mortality. 
We explain this discrepancy that
in the case of a large world graph the infection rates in the network are deviating from the simple mass-action-law forms (\ref{infections_rate}). 
The validity of mass-action laws for the infection rates requires that the compartmental populations are fully mixed
and spatially homogeneous.
The deviation from mass-action law can be explained by the observation that in large world graphs it needs a certain delay time until the compartmental populations are sufficiently well mixed such that the local infection rates become homogeneous in the whole network and independent from position nodes of the walkers. Therefore, the infection rates take mass-action laws asymptotically in the large-time limit. This retardation effect can be clearly identified in Fig. \ref{WS_mort_nomort}, see especially left frames without mortality.

In the simulation runs of Fig. \ref{BA_mort_nomort_conf} we investigate the impact of confinement measures in absence of mortality. We mimic the effect of walkers subjected to confinement
by varying the transmission probability $p_n$ that an I-walker infects an S-node.
We interpret the quantity $1-p_n$ as the fraction of confined walkers which are disabled to transmit the disease. $p_n=1$ corresponds to no confinement and $p_n=0$ to full confinement of the walkers' population.
The lower right frame of Fig. \ref{BA_mort_nomort_conf} shows that
for both types of graphs (BA and WS) $R_0$ falls off linearly with respect to the fraction $1-p_n$ of confined walkers.
This reflects a linear relationship between $p_n$ and $R_0$. For the same fractions of confined walkers,
the $R_0$ values of the small-world BA graph are larger than those of the large-world WS graph. This confirms the feature of high susceptibility of small world networks to epidemic spreading.
The decrease of $R_0$ with the fraction of confined walkers also gives strong evidence for the efficiency and need of confinement measures, especially for populations living in small world networks such as agglomerations and cities.

\section{Conclusions}
\label{Conclusions}

We established a stochastic mean field compartment model for vector transmitted diseases, where we take into account mortality of the individuals during the phase of the disease. Based on the assumption that vectors do not fall ill, we neglected mortality of the vectors. Moreover, we neglect demographic effects coming from natural natality and mortality. Our model allows for interpretations beyond epidemiological dynamics, for instance has an analogy to certain chemical reactions \cite{Simon2020}, propagation of wood fires, pollutants, and others.

Comparison of the results of the mean field model with random walk simulations shows that the mean field model
assuming simple forms of the infection rates (Eqs. (\ref{infections_rate})) are well capturing the spreading dynamics in small world networks and vanishing mortality (see Figs. \ref{BA_mort_nomort}, \ref{BA_mort_nomort_conf}). The agreement is less well in large world networks (Fig. \ref{WS_mort_nomort}) in the beginning of the epidemic, but again agrees well for large observation times in absence of mortality.
Larger deviations between mean field and random walk occur for higher mortality. 
For future research, a more sophisticated mean field model would be desirable, which is able to better capture topological effects of large world (weakly connected) networks. 
A promising direction to advance with this task is to extend our model to a
degree-based mean-field approach \cite{Kyriakopoulos-et-al2018}. In this way topological information of the network is incorporated which might avoid the drawbacks of mass-action law infection rates for small observation times.

Another direction of interest is the spreading with modified steps in the random walks, such as for instance under resetting, mimicking the influence of long-range journeys of individuals
\cite{resetting_first,mi_resetting_chaos2025,ria_resetting_chaos2025,Pal_Sandev2015,Trajanovski2025}.
Further promising extensions of the present model are opened by non-monotonous infection rates (different from simple mass-action-laws) for which under certain conditions the endemic equilibrium exhibits bifurcations, allowing for emergence of chaotic attractors \cite{bes_mi_chaos2025}. 
Also the inclusion of demographic effects originating from natural natality and mortality may open an interesting generalization of the present approach.

\end{document}